\documentclass[aps,prd,reprint,nofootinbib,longbibliography]{revtex4-1}
\usepackage{blindtext}
\usepackage{enumitem}
\usepackage{mathtools}
\usepackage{xcolor}
\usepackage{adjustbox}
\usepackage{graphicx}
\usepackage{cancel}
\usepackage{color, colortbl}
\usepackage[first=0,last=9]{lcg}
\definecolor{Gray}{gray}{0.9}
\usepackage{lipsum, babel}
\usepackage{bm}
\usepackage[utf8]{inputenc}
\usepackage{float}
\usepackage{hyperref}
\usepackage{ulem}
\usepackage[capitalize]{cleveref}
\usepackage{titlesec}
\hypersetup{
    colorlinks=true,
    linkcolor=blue,
    citecolor=red,
    filecolor=magenta,      
    urlcolor=magenta,
}

\newcommand\funop[1]{\mathop{{}#1}}
\newcommand{\dd}{\mathop{}\!\mathrm{d}}
\newcommand{\mrm}[1]{\mathrm{#1}}
\newcommand{\usim}{\mathord{\sim}}
\newcommand{\bv}[1]{\mathbf{#1}}

\newcommand{\uni}[1]{\, \mathrm{#1}}
\newcommand{\mcal}[1]{\mathcal{#1}}

\usepackage{makeidx}
\makeindex

\begin{document}

\title{\boldmath Multi-messenger signal from QCD-Axion Ultracompact Minihalos}
\author{Dorian W.~P.~Amaral$^{1}$, Enrico D. Schiappacasse$^{2*}$, Karla Tapia-Rebolledo$^{3}$}

\affiliation{
$^1$ Institut de F\`isica d’Altes Energies (IFAE), The Barcelona Institute of Science and Technology,
Campus UAB, 08193 Bellaterra (Barcelona), Spain\\
$^2$ Facultad de Ingenier\'ia, Universidad San Sebasti\'an, Bellavista 7, Santiago 8420524, Chile\\
$^3$ Departamento de Ciencias F\'isicas, Universidad Andr\'es Bello, República 220, Santiago 8370134, Chile}

\begin{abstract}
We explore the multimessenger signatures from ultracompact minihalos (UCMHs) seeded by stellar-mass primordial black holes (PBHs) and composed of QCD-axion dark matter. Using the secondary-infall framework, with relic-abundance and isocurvature conditions from a pre-inflationary Peccei--Quinn scenario, we numerically recover the characteristic halo profile and adopt a conservative constant-density core for the innermost region. We study the gravitational wave and electromagnetic signatures from neutron-star--PBH inspirals inside these dense structures. Dynamical friction produces a gravitational-wave dephasing relative to the vacuum binary, observable with LISA for primordial black holes with masses $\lesssim\! 10M_\odot$. Simultaneously, gravitationally focused axions can resonantly convert into photons in the neutron-star magnetosphere, producing a narrow-band radio signature. For nearby galactic neutron stars, the projected sensitivity of the Green Bank radio telescope GBT captures the QCD-axion band in the \(\mu{\rm eV}\) mass range. Gravitational waves therefore probe the presence and structure of the UCMH, while the radio signal probes the particle nature of its dark matter, making their joint observation a distinctive test of a mixed dark sector.
\end{abstract}
\maketitle

\section{Introduction}
\label{sec:introduction}

The nature of dark matter (DM) remains one of the central open questions in modern physics. A broad body of astrophysical and cosmological observations provides compelling evidence for a cold, non-luminous matter component, from galactic to the cosmic microwave background and the formation of large-scale structure. Nevertheless, despite its remarkable success as a phenomenological ingredient in the standard cosmological model, the microscopic origin of DM remains unknown\,\cite{Bertone:2018krk,Bertone:2016nfn, Bertone:2004pz}.

Motivated by unresolved problems in the Standard Model of particle physics, the QCD-axion stands out as one of the most compelling dark matter candidates\,\cite{PhysRevLett.38.1440, PhysRevLett.40.223, PhysRevLett.40.279, PRESKILL1983127, ABBOTT1983133,DINE1983137}. The search for axion dark matter has gained remarkable momentum in recent years, giving rise to a diverse experimental and observational program that spans haloscopes\,\cite{Hagmann:1990tj,2001PhRvD..64i2003A,ADMX:2018gho,HAYSTAC:2018rwy}, helioscopes\,\cite{CAST:2017uph}, and indirect astrophysical probes\,\cite{Visinelli:2024tyw,Long:2024qvd,Escudero:2023vgv,Choi:2022btl, Schiappacasse:2021zlr, Fujikura:2021omw,Nurmi:2021xds,Hertzberg:2020dbk,Hertzberg:2020hsz,Leroy:2019ghm,Huang:2018lxq,Hook:2018iia,Safdi:2018oeu,Iwazaki:2017rtb,Iwazaki:2014wka}. Many of these efforts exploit the axion--photon interaction, $g_{\phi\gamma\gamma}\sim \phi\,{\bf E}\cdot{\bf B}$, 
where $\phi$ is the axion field, $g_{\phi\gamma\gamma}$ is the axion--photon coupling, and {\bf E} and {\bf B} denote the electromagnetic fields. This coupling is particularly compelling because it allows axions to convert into photons in the presence of external magnetic fields, providing one of the central experimental handles in the search for axion dark matter. 

On the other hand, primordial black holes (PBHs) provide a qualitatively different and well-motivated dark matter candidate. Depending on their mass and abundance, PBHs could account for either the entire DM density\,\cite{Inomata:2017okj, Kawasaki:2016pql}  or a fraction of it\,\cite{Carr:2025kdk, Carr:2021bzv, Carr:2020gox}. In particular, gravitational-wave observations by LIGO--Virgo-KAGRA\,\cite{KAGRA:2021duu, Feng:2024obn,Sasaki:2016jop,LIGOScientific:2016aoc} and the recent NANOGrav results\,\cite{Kohri:2020qqd,NANOGrav:2020bcs,Inomata:2020xad}  have renewed interest in the possibility that a small fraction of DM may reside in PBHs with masses around $(1-10)\,M_\odot$. Primordial Black Holes are among the earliest proposed DM candidates\,\cite{Hawking:1971ei,Hawking:2026,1975Natur.253..251C}, and their phenomenology has been extensively investigated both as a standalone component and in mixed-DM scenarios involving other candidates, including axions\,\cite{Yin:2024xov,Hertzberg:2020hsz,Ferrer:2018uiu} and weakly interacting massive particles\,\cite{Hertzberg:2020kpm,Adamek:2019gns,Scott:2009tu}.

Multimessenger signatures involving QCD axions and compact objects
have previously been investigated for dark-matter overdensities
surrounding intermediate-mass black holes\,\cite{Edwards:2019tzf}. Under the assumption that the central black hole grows adiabatically, the surrounding dark-matter distribution is expected to develop a steep spike that can be parameterized as $\rho_\text{DM}\propto (r_\text{sp}/r)^{\gamma_\text{sp}}$  
with 
$2r_\text{sch} < r < r_\text{sp}$, where 
$r_\text{sch}$ denotes the Schwarzschild radius,
$r_\text{sp}$ is the characteristic spike radius. The spike index $\gamma_\text{sp}$ is typically expected to lie within the range
$\gamma_\text{sp} \in [9/4,5/2]$\,\cite{Gondolo:1999ef}. The emergence of such a steep distribution is therefore contingent on an idealized formation history, particularly on sufficiently slow black-hole growth at, or very near, the halo center. More realistic growth histories can substantially weaken the enhancement; for instance, rapid black-hole formation or the formation of a massive seed away from the dynamical center can produce a considerably shallower profile~\cite{Ullio:2001fb}. The spike may subsequently be further eroded by black-hole and host-halo mergers~\cite{Merritt:2002vj} or by gravitational scattering from stars in a dense nuclear environment, which tends to drive the dark-matter distribution toward a shallower cusp with  $\gamma_\mathrm{sp} = 3/2$~\cite{Gnedin:2003rj,Merritt:2003qk}.

These limitations do not directly apply to the formation mechanism considered here. In a PBH-seeded ultracompact minihalo (UCMH)~\cite{Ricotti:2009bs}, the central black hole forms in the early Universe before the surrounding dark-matter structure is assembled. The PBH subsequently acts as a pre-existing gravitational seed for cosmological dark-matter accretion, rather than being introduced through the slow growth of an astrophysical seed inside an equilibrium halo. Consequently, the centrality and adiabatic-growth assumptions required by the conventional black-hole spike scenario are absent. Numerical simulations for generic dark matter have found that steep dark-matter distributions with a power-law density profile $\rho_\mathrm{halo}(r) \propto r^{-9/4}$ can indeed develop around PBHs, in agreement with the basic secondary-infall picture~\cite{Ricotti:2009bs,Adamek:2019gns}. Their subsequent survival may still depend on tidal interactions and on the later astrophysical environment~\cite{Nurmi:2021xds,Hertzberg:2019exb}, but the formation of the steep profile does not rely on the highly restrictive black-hole growth history underlying conventional adiabatic spikes.

The mixed-DM framework explored in this work, where pre-inflationary QCD axions make up the dominant DM component and PBHs provide a subdominant contribution, naturally predicts the existence of PBHs surrounded by axion minihalos in the Milky Way today. Physically, the PBH behaves as a compact gravitational seed that draws in the ambient axion DM. According to the standard theory of spherical gravitational collapse \,\cite{1985ApJS...58...39B}, localized overdensities in the DM distribution can grow into gravitationally bound minihalos. In this scenario, the smooth axion background is progressively accumulated around PBHs, mainly after matter-radiation equality, leading to the formation of ultracompact minihalos (UCMHs) that primarily survive as part of present-day galactic halos.

Ultracompact minihalos can ultimately collide 
with highly magnetized objects, such as neutron stars (NSs), opening up the possibility for generating multimessenger signals.
Accounting for tidal disruption events, the
differential NS-UCMH encounter rate as a function of the galactocentric radius is approximately $\dd \Gamma_\text{NS-UCMH}/\dd r \sim 10^{-7}\,\text{kpc}^{-1}\,\text{s}^{-1}$, leading to a total galactic encounter rate of the order of $0.1\,\mrm{day^{-1}}$ for PBH masses $(1\text{--}10) M_\odot$\,\cite{Nurmi:2021xds}.
Following a capture, the NS-PBH system can undergo an inspiral, generating a multimessenger signal consisting of gravitational and electromagnetic radiation. 

On the gravitational side, the inspiral of a neutron star around a central PBH with mass $(1-10)M_\odot$ naturally produces low-frequency gravitational waves (GWs) that are detectable with the planned space-based interferometer LISA\,\cite{LISA2017}. On the electromagnetic side, QCD axions from the surrounding minihalo resonantly convert into photons as they cross the neutron-star magnetosphere, generating a narrow radio signal set primarily by the axion mass. For axion masses $10^{-6}\text{--}10^{-4}\,{\rm eV}$, the produced radiation has frequencies of MHz up to tens of GHz---a window covered by existing radio telescopes. In this range, the electromagnetic counterpart can be searched for as a narrowband radio line. This mass range for the axion is therefore especially interesting: it connects radio searches for axion--photon conversion with low-frequency gravitational-wave observations, offering a complementary way to probe QCD-axion minihalos around stellar-mass PBHs.

In this paper, we propose a new pathway for detecting the QCD axion through multi-messenger astrophysics \cite{Meszaros:2019xej} via UCMHs. We apply the standard secondary-infall framework to QCD axion dark matter, incorporating the relic-abundance and isocurvature constraints characteristic of a pre-inflationary Peccei–Quinn symmetry-breaking scenario. We reproduce the expected power-law density profile and dark-matter concentration found in previous studies for generic particle dark matter\,\cite{Berezinsky:2013fxa,Ricotti:2007au}. This outer  envelope fixes the matching density at the angular-momentum-supported inner region\,\cite{2012PhRvD..86d3519L, Ricotti:2009bs}. Our results show that the accretion process is governed predominantly by the gravitational field of the central PBH, with axion isocurvature perturbations and the initial velocities of the accreting matter shells producing only negligible corrections.

 When such systems undergo close encounters with highly magnetized galactic objects, such as neutron stars, they can generate a characteristic multimessenger signal during the NS--PBH inspiral. The gravitational-wave channel can therefore reveal the presence and gravitational influence of the UCMH through a dephasing relative to the vacuum compact-binary waveform. Simultaneously, axions from the same minihalo can resonantly convert into photons in the neutron-star magnetosphere, producing a narrow-band radio counterpart that directly probes the axion particle component. Our projected sensitivities show that the minimum detectable axion--photon coupling can lie well below the canonical QCD-axion coupling in the relevant mass range for typical galactic NSs. Thus, the combined detection of gravitational-wave dephasing and radio axion-conversion emission would provide a distinctive signature of a PBH-seeded axion UCMH, testing compact dark matter, its surrounding dark-matter distribution, and the particle nature of dark matter within a single astrophysical event.

\section{QCD-Axion ultracompact minihalos}

Primordial black holes can seed ultracompact minihalos, acting as
localized isocurvature seeds in the dark-matter background and driving
the accretion of the surrounding dark matter
\cite{Berezinsky:2013fxa}. The range \(M_{\rm PBH}\sim(1\text{--}10)M_\odot\) is particularly motivated because the horizon mass during the QCD confinement epoch is of order \(M_\odot\). Around this transition, the effective sound speed of the cosmic plasma is reduced, weakening pressure support against collapse and enhancing PBH formation near the corresponding horizon mass scale \cite{Jedamzik1997,Musco2024}. However, current data do not allow such objects to make up an ${\cal O}(1)$ fraction of the dark matter.
Constraints from LIGO/Virgo on PBH binary mergers, explicitly including the effects of local DM minihalos, imply
$f_{\rm PBH}\lesssim 10^{-3}$ for
$10\,M_\odot\lesssim M_{\rm PBH}\lesssim300\,M_\odot$
\,\cite{Kavanagh:2018ggo}.
CMB bounds accounting for the formation of a DM halo around each PBH also constrain their abundance.
For the disk-accretion scenario, which yields the stronger CMB limits, one finds approximately
$f_{\rm PBH}\lesssim 2\times10^{-2}$ at
$M_{\rm PBH}\simeq10\,M_\odot$, with the constraint becoming weaker toward lower PBH masses
\,\cite{Serpico:2020ehh}.
Additionally, microlensing observations such as Icarus imply approximately
$f_{\rm PBH}\lesssim10^{-1}$ in the stellar-mass range
\,\cite{Oguri:2017ock}.
These results motivate the regime studied in this work: a predominantly QCD axion dark sector containing a rare population of stellar-mass PBHs, sufficiently sparse to satisfy current bounds but sufficiently massive to seed dense ultracompact minihalos and generate distinctive multi-messenger signatures.

PBHs act as localized gravitational seeds embedded in the particle-dark-matter background, which inevitably leads to the growth of spherically symmetric minihalos according to the theory of spherical gravitational collapse or secondary infall~\cite{1985ApJS...58...39B}. Around the time of radiation-matter equality, self-similar dark matter halos begin to grow around these seeds, with masses and radii that are dependent on redshift $z$. For the case of an initial small fraction of dark matter in bare PBHs without a UCMH, $f_{\text{PBH}} \ll 1$, with typical masses $M_\mathrm{PBH}$,  these compact objects may be considered as an isolated seed. This leads to UCMHs that grow with redshift as $M_\mathrm{UCMH} \propto M_\mathrm{PBH}/(1+z)$, having a mass-radius relation $R_\mathrm{UCMH} \propto M_\mathrm{UCMH}^{1/3}/(z+1)$\,\cite{Mack:2006gz, Berezinsky:2013fxa, Ricotti:2009bs}. The growth of UCMHs halts when they begin to interact with non-linear structures at around $(z \sim 30-10)$~\cite{Berezinsky:2013fxa}, so that the UCMH abundance is given by $f_{\text{UCMH}}\sim 10^2 f_{\text{PBH}}$. 

Ultracompact minihalos are appealing astrophysical objects in the dark matter framework due to their steep radial density profile $\rho_\mathrm{UCMH}(r) \propto r^{-9/4}$~\cite{Serpico:2020ehh, Adamek:2019gns}. This steep profile allows the dark-matter density in the innermost regions of UCMHs to exceed the smooth local galactic density by several orders of magnitude. This enhances direct detection prospects through dense tidal streams~\cite{Choi:2022btl} and favors axion-star nucleation, thereby opening additional avenues for indirect searches~\cite{Yin:2024xov,Hertzberg:2020dbk}.

\section{QCD axion dark matter}
\label{dominant}

We take the QCD axion to constitute the dominant particle dark-matter component surrounding a subdominant population of PBHs. We focus on the pre-inflationary Peccei--Quinn (PQ) scenario, where the PQ symmetry is broken before or during inflation and is not restored later. In this scenario, the axion field is homogenized over our observable Universe, and its current abundance is generated mainly by the vacuum-misalignment mechanism.  Here, we focus only on the ingredients relevant to PBH-seeded UCMH formation: the axion relic abundance and the primordial isocurvature spectrum. Standard formulae and conventions are reviewed in Ref.~\cite{Marsh:2015xka}.

For the QCD axion, the zero-temperature axion mass and axion decay constant are related
by~\cite{GrillidiCortona:2015jxo}
\begin{equation}
m_a \simeq 5.70 \mu\text{eV}\,\left( \frac{10^{12}\,\text{GeV}}{F_a} \right)\,,
\end{equation}
where $F_a \equiv f_a/N_{\text{DW}}$ is the physical axion decay constant with $N_\text{DW}$ the domain wall number. For the benchmark value used in our numerical analysis, we have 
$m_a = 2.0 \times 10^{-5}\,\text{eV}$, which corresponds to $F_a \simeq 2.85 \times 10^{11}\,\text{GeV}$. 

In the axion mass regime of interest, $10^{-6}\,\text{eV} \lesssim m_a \lesssim 10^{-4}\,\text{eV}$, the QCD axion begins oscillating before the QCD phase transition. Therefore, we use the standard misalignment result:\,\cite{Fox:2004kb, Marsh:2015xka}
\begin{widetext}
\begin{equation}
\Omega_{a}h^2\sim 2 \times 10^4 \left(  \frac{F_a}{10^{16}\,\text{GeV}} \right)^{7/6}\left[\theta^2_{a,i} + H^2_I/(2\pi F_a)^2\right]\funop{\mathcal{F}_\text{anh}\!\left(\sqrt{\theta^2_{a,i} + H^2_I/(2\pi F_a)^2}\right)}\,,
\label{eq:DM}
\end{equation}
\end{widetext}
where $\theta_{a,i}$ is the initial misalignment angle,   and $\mathcal{F}_\text{anh}(x) \equiv \{\text{ln}[e/(1-x^2/\pi^2)]\}^{7/6}$ is a correction factor related to QCD potential anharmonic effects. For the benchmark values considered here, the inflationary contribution driven by $H_I/F_a$ is negligible compared with the initial misalignment angle, and the relic abundance is therefore mainly controlled by \(\theta_{a,i}\). Since $f_\text{PBH} \ll 1$, we require the axion to reproduce essentially all of the observed cold-dark-matter abundance, so that $\Omega_a h^2 \simeq 0.120$\, \cite{Planck:2018vyg}.

 In addition to the homogeneous relic abundance, a light axion field during inflation acquires quantum fluctuations $\delta a \simeq H_I/(2\pi)$, which generate an isocurvature component statistically uncorrelated with the adiabatic curvature perturbation. Including anharmonic corrections,  the dark-matter-normalized isocurvature power spectrum is
\begin{equation}
\Delta^2_{{\rm iso},a}(k)
=
(f_a^{\rm DM})^2\mathcal{A}^2_{\text{anh}}
\frac{H_I^2}{\pi^2 \left(F_a\theta_{a,i}\right)^2}
\left(\frac{k}{k_0}\right)^{n_{\rm iso}-1},
\label{eq:qcdaiso_pivot}
\end{equation}
where $f_a^{\rm DM}\equiv \Omega_a/\Omega_{\rm DM}\simeq 1$ and $\mathcal{A}_{\text{anh}}\equiv 1+(\theta_{a,i}/2)\,\dd(\text{ln} \mathcal{F}_{\text{anh}})/\dd\theta_{a,i}$. 

The primordial axion isocurvature amplitude is constrained by CMB observations. For an uncorrelated CDM isocurvature mode with $n_\text{iso}=1$, Planck 2018 gives\,\cite{Planck:2018jri}  
\begin{equation}
\beta_\text{iso}(k_0) \equiv
\frac{\Delta^2_{{\rm iso},a}(k_0)}
{\Delta^2_{\mathcal R}(k_0)+\Delta^2_{{\rm iso},a}(k_0)}\, < \,0.038
\end{equation}
 at $k_0 = 0.05\, \text{Mpc}^{-1}$ (95$\%$ CL). Using  \(A_s \equiv \Delta^2_{\mathcal R}(k_0)=2.10\times10^{-9}\) , the saturated Planck bound corresponds to\,\cite{Planck:2018jri} 
\begin{equation}
\Delta^2_{{\rm iso},a}(k_0)\simeq8.3\times10^{-11}.
\label{eq:qcdaiso_bound}
\end{equation}
We adopt this limiting value in our benchmark calculation to maximize the allowed primordial axion-isocurvature contribution while remaining consistent with CMB constraints.

Substituting these benchmark values into Eq.\,\eqref{eq:qcdaiso_pivot}, we have 
$H_I \simeq 7.46 \times 10^6\,\text{GeV}$. Note that this inflationary Hubble scale is not an additional assumption, but follows from simultaneously requiring  the QCD axion to constitute the dark matter and to saturate the Planck isocurvature limit for our chosen axion mass benchmark.

For completeness, we assume a conventional single-field slow-roll inflationary sector sourcing the observed adiabatic perturbations. During inflation, the light axion is only a spectator field and does not generate the dominant adiabatic spectrum. The inflationary Hubble scale is related to the tensor-to-scalar ratio through $r=(2H_I^2)/(\pi^2M_{\rm Pl}^2A_s)$\,\cite{Baumann:2009ds}. For the benchmark values, we have $r\simeq 9.5 \times 10^{-16}$, so that $n_{\text{iso}} \simeq 1 - r/8$ is essentially unity and the assumption \(n_{\text{iso}}=1\) entering the Planck isocurvature constraint is self-consistent.

\section{Spherical accretion model}
\label{sec.accretionmodel}

\subsection{Numerical QCD-axion UCMH density profile} 

We consider a scenario in which the majority of dark matter is composed of QCD-axion particles alongside a smaller fraction of PBHs with masses  \(M_\mrm{PBH} \sim(1\text{--}10)\, M_{\odot}\). With an initial PBH fraction of \(f_{\text{PBH}} \lesssim 10^{-3}\), the subset of these PBHs that do not form binaries in the early Universe is approximately \( \usim \text{exp}(-f_{\text{PBH}}) \gtrsim 0.999\) (see Appendix~\ref{app:fractionisolated}), and therefore represents almost the entire PBH population. We focus on these isolated PBHs, since close encounters between PBHs in early binaries can disrupt and unbind their surrounding minihalos\,\cite{Kavanagh:2018ggo}. Later on in the history of the Universe, these same PBHs can form binary systems with neutron stars and generate the multimessenger signatures we look for.

Primordial black holes gravitationally accrete the surrounding axion dark matter. In what follows, we numerically reconstruct the local axion-dark-matter density profile by extending the standard spherical secondary-infall framework \cite{Gunn:1972sv,Fillmore:1984wk, 1985ApJS...58...39B} to a PBH-seeded minihalo \cite{Mack:2006gz,Berezinsky:2013fxa}. In contrast to generic pressureless-dark-matter treatments, our initial conditions explicitly include the QCD axion isocurvature perturbations enclosed within each accreting shell. We present here the essential ingredients, physical interpretation, and overall flow of the calculation, while the detailed derivations, intermediate steps, and numerical implementation are provided in Appendix~\ref{app:accretion_details}.

In the framework of our multimessenger analysis, we are primarily interested in QCD axion masses in the range $10^{-6} \lesssim m_a \lesssim 10^{-4}\,\text{eV}$ accreting around isolated and 
stationary PBHs with $M_\text{PBH} \sim (1-10)\,M_\odot$. In this parameter range, coherent axion oscillations begin before the PBH forms. Consequently, once the PBH is in place, it is already embedded in an axion component that behaves as cold dark matter and is available for accretion.
Within the spherical secondary-infall framework, we follow the evolution of concentric matter shells. Each shell initially expands approximately with the cosmological background, subsequently decouples from the Hubble flow, and eventually collapses and virializes. We neglect the dynamical distinction between baryons and axion dark matter during halo formation and describe the accreting material as a single effective collisionless component. The dark-matter fraction $f_\chi = \Omega_\text{DM}/\Omega_m$
is restored when normalizing the axion-halo density. A fully coupled treatment of axion and baryonic components, including the delayed infall of baryons after recombination, is beyond the scope of the present analysis.
We adopt the conventional approximation where the virial radius is given by the half of the turnaround radius\,\cite{CooraySheth2002}. 

For each shell, we choose an initialization time \(t_i\) after both the PBH has formed and the axion field has begun to behave as cold dark matter: $t_i \gtrsim \text{max}(t_\text{PBH},t_\text{osc})$, where $t_{\rm PBH}$ is the PBH formation time and $t_{\rm osc}$ is the onset time of coherent axion oscillations. We also require the smooth axion perturbation to remain sufficiently small for the linear prescription used to initialize the shell motion to be valid.

The shell dynamics are conveniently described in terms of the departure from the Hubble flow, $r(t)=a(t)\,b(t)\,\xi$,
where $a(t)$ is the cosmological scale factor,  $\xi$ is the comoving shell radius, and $b(t)$ measures the departure of the shell from the unperturbed Hubble expansion. Using a matter-radiation background and parameterizing time by $
y \equiv a(\eta)/a_{\rm eq}$, the shell equation can be written as
\begin{equation}
\begin{split}
&(1+y)y\,b''(y)+\left(1+\frac{3}{2}y\right)b'(y)\\
&-\frac{b(y)}{2}
+\frac{(1+\delta_i)b_i^3}{2b^2(y)}=0\,.
\end{split}
\label{eq:secIV_main_shell}
\end{equation}
Here \(b_i\equiv b(y_i)\) denotes the value of the shell-deformation function at the initial time, with $y_i = a(\eta_i)/a_\text{eq}$. We set \(b_i=1\), which amounts to choosing the comoving shell coordinate \(\xi\) such that the initial physical radius is \(r(y_i)=a(y_i)\xi\).

The initial overdensity of each shell receives contributions from the central PBH and from the axion inhomogeneity:
\begin{equation}
\delta_i
=
\delta_{\rm PBH}(r_i)+\delta_{{\rm halo},i}\,.
\end{equation}
For the isocurvature contribution, we adopt a representative positive rms fluctuation, $\delta_{{\rm halo},i} = +\sigma_\text{iso,i}$.
This explicitly incorporates the axion isocurvature perturbation enclosed within each shell, even though its ensemble average vanishes.

 Smoothing the isocurvature perturbation over the comoving shell scale $R=\xi$, its variance at initialization is
\begin{equation}
\sigma^2_{{\rm iso},i}
=
\left(1+\frac{3}{2}y_i\right)^2
\Delta^2_{\rm iso,a}(k_0)
 (Rk_0)^{(1-n_{\rm iso})}
I(n_{\rm iso})\,.
\label{eq:secIV_sigma_iso}
\end{equation}
Here, $f_a^{\rm DM} \simeq 1$, $I(n_{\rm iso})$ denotes the dimensionless window-function integral and $k_0$ is the pivot scale. The factor $(1+3y_i/2)$ accounts for the linear evolution of the isocurvature perturbation up to the initialization time. Its subsequent nonlinear evolution is determined by the shell equation, \cref{eq:secIV_main_shell}.

The initial shell velocity is fixed by linearizing mass conservation at the starting time. In terms of $y$, this gives
\begin{equation}
\left.\frac{\dd b}{\dd y}\right|_{y_i}
=
-\frac{1}{3(1+\sigma_{{\rm iso},i})}
\left.\frac{\dd \sigma_{\rm iso}}{\dd y}\right|_{y_i}\,,
\label{eq:app_bprime_sigma}
\end{equation}
where
\begin{equation}
\left.\frac{\dd \sigma_{\rm iso}}{\dd y}\right|_{y_i}
=
\frac{3}{2}\Delta_{\rm iso,a}(k_0)
(Rk_0)^{(1-n_{\rm iso})/2}
I^{1/2}(n_{\rm iso})\,.
\label{eq:secIV_sigma_prime}
\end{equation}
Together with the PBH contribution to $\delta_i$, Eqs.~\eqref{eq:secIV_main_shell}--\eqref{eq:secIV_sigma_prime} fully determine the initial-value problem for each shell.

\begin{figure}[t]
\centering
\includegraphics[width=8 cm]{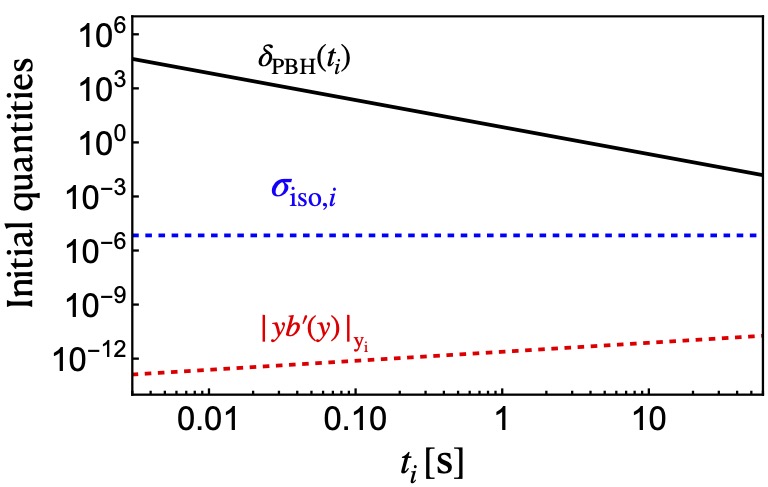}
\includegraphics[width=8 cm]{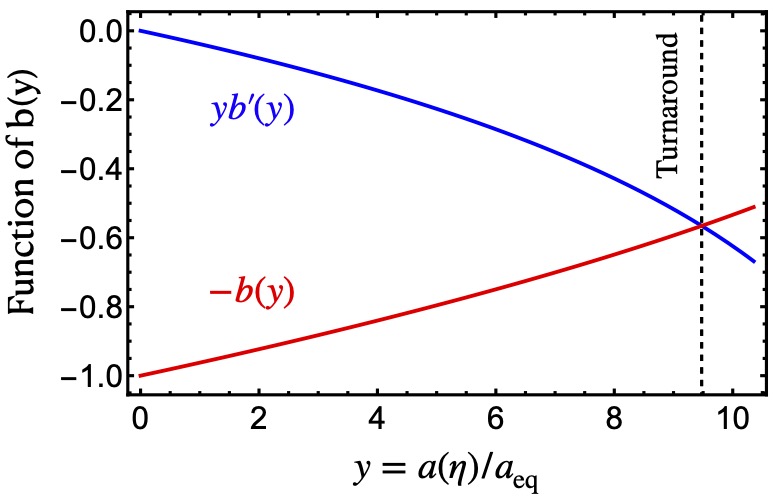}
\includegraphics[width=8 cm]{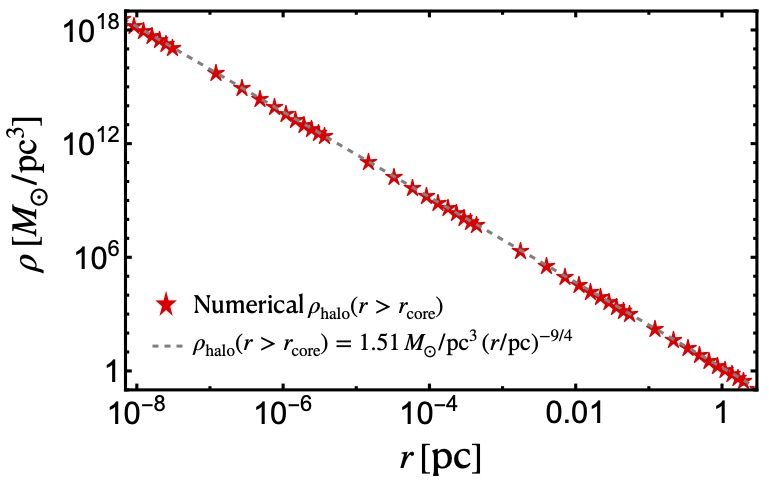}
\caption{\textbf{Top:} Quantities at initialization, including the overdensity from the central PBH $\delta_\mrm{PBH}$, axion isocurvature variance $\sigma_{\mrm{iso}, i}$, and fractional departure from Hubble expansion $|y_ib'_i|$. \textbf{Middle:} Turn-around radius calculation 
for the shell with initial time $t_i \simeq 21\,\mrm{s}$. \textbf{Bottom:} Numerical QCD-axion minihalo density profile.}
\label{fig.ucmhperfil}
\end{figure}

We then evolve the shells until turnaround, defined by $\dot r=0$, or equivalently
\begin{equation}
b(y_{\max})+y_{\max}b'(y_{\max})=0\,,
\end{equation} 
where \(y_{\max}\) denotes the value of \(y\) at turnaround, and we define $b_{\max}\equiv b(y_{\max})$. The maximum physical shell radius is therefore $r_{\max}=a_{\rm eq}y_{\max}b_{\max}\xi$.
We adopt the conventional approximation that the virialized halo radius is one-half of the turnaround radius, $R_\text{halo}\equiv r_\text{max}/2$ \,\cite{CooraySheth2002}.
Neglecting shell crossing, the matter initially enclosed within \(r_i\) remains enclosed within the final radius \(R_{\rm halo}\), such that $M_{\rm halo,i}(<r_i)=M_{\rm halo}(<R_{\rm halo})$. The mean enclosed axion density is then
\begin{equation}
\begin{split}
\bar\rho_{\rm halo}(<R_{\rm halo})
 &=
\frac{6f_\chi M_{{\rm halo},i}(<r_i)}{\pi r_{\rm max}^3}\\
& =
\frac{f_\chi\rho_{\rm eq}(1+\sigma_{{\rm iso},i})}{y_{\max}^3 b_{\max}^3/4}\,,
\label{eq:app_rhohalo}
\end{split}
\end{equation}
where $\rho_{\rm eq}$ is the background matter density at matter-radiation equality. For a spherical halo, the local axion density at \(R_{\rm halo}\) is obtained from the radial variation of the enclosed halo mass. Assuming that the mean enclosed density follows \(\bar\rho_{\rm halo}(<R_{\rm halo})\propto R_{\rm halo}^{-\gamma}\), the local and mean enclosed densities are related by
\begin{equation}
\begin{split}
\rho_{\rm halo}(R_{\rm halo}) &= \frac{1}{4\pi R^2_{\rm halo}}\frac{\dd M_{{\rm halo},i}(<R_{\rm halo})}{\dd R_{\rm halo}} \\
&=\left(1-\frac{\gamma}{3}  \right)\bar\rho_{\rm halo}(<R_{\rm halo})\,.
\label{eq:rhonum}
\end{split}
\end{equation}
Figure\,\ref{fig.ucmhperfil} shows numerical results for the  case $M_\text{PBH} = 1 M_\odot$, $f_\text{PBH} = 10^{-3}$, $m_a \simeq 20\,\mrm{\mu eV}$, $n_\text{iso}=1$, and $\Delta^2_\text{iso,a}(k_0) = 8.3\times 10^{-11}$, $\theta_{a,i} \simeq 1.03$, $H_I \simeq 7.46 \times 10^6 \,\text{GeV}$, $F_a \simeq 2.85 \times 10^{11} \,\text{GeV}$,
and $f^a_\text{DM} \simeq 1$. We anticipate that in
the mass parameter space for the QCD axion to study radio emissions during the inspiral of a neutron star around the central PBH of a UCMH, the mass range of interest is \(10^{-6}\,\mathrm{eV}\lesssim m_a \lesssim 10^{-4}\,\mathrm{eV}\). In this range, \(t_{\rm osc} \ll t_{\rm PBH}\), implying that the central PBH starts accreting a minihalo essentially immediately after its formation. 

The top panel shows the initial conditions for each shell at different initialization times,  $\delta_{\rm PBH}(r_i)$, the QCD-axion isocurvature rms fluctuation, $\sigma_{\mathrm{iso},i}$, and the fractional departure from pure Hubble expansion, $|y_ib_i'|$. We consider the range $0.003\,\text{s} \lesssim t_i \lesssim 60\,\text{s}$, corresponding approximately to shells that populate the envelope outside the angular-momentum-supported core (see the next subsection) up to shells whose virialized radii are comparable to the conventional UCMH size near the epoch when nonlinear structures begin to form.
In radiation domination, the initial quantities show different time dependencies: 
\begin{equation}
\delta_{\rm PBH}(r_i) \propto t_i^{\rm -3/2}, ~~ \sigma_{\mathrm{iso},i}\simeq \text{constant}, ~~  |y_ib_i'| \propto t_i^{1/2}\,. 
\end{equation}
The hierarchy over the range considered here is given by $|y_ib_i'| \ll \sigma_{\mathrm{iso},i} \ll \delta_{\rm PBH}(r_i)$, so that
 the PBH controls the initial overdensity, the axion isocurvature provides a small nearly scale-independent correction, and the initial peculiar motion away from the Hubble flow is tiny.
 
The middle panel shows the two functions $-b(y)$ and $yb'(y)$ for a shell initialized at $t_i \simeq 21\,\mrm{s}$. Their intersection determines  $y_{\rm{max}}$ since the turnaround condition is  $b(y_{\mathrm{max}}) + y_{\mathrm{max}} b'(y_{\rm{max}})=0$. The corresponding turnaround radius is then obtained from $r_{\rm{max}} = a_{\rm{eq}}y_{\rm{max}}b(y_{\rm{max}})\xi$. 

The bottom panel shows the numerically reconstructed density profile of the QCD-axion minihalo. Fixing the theoretically expected slope to \(-9/4\), the numerical profile is well fitted by
\begin{equation}
\rho_{\rm halo} \simeq 1.02 \times 10^{-22}\,\text{g cm}^{-3}
\left( \frac{r}{\text{pc}} \right)^{-9/4}
\left( \frac{M_{\rm PBH}}{M_\odot} \right)^{3/4}\,.
\label{eq:rho94num}
\end{equation}
Allowing the slope to vary independently gives \(\gamma=2.244\), in excellent agreement with the characteristic value \(9/4\).
Repeating the numerical reconstruction for \(M_{\rm PBH}=10\,M_\odot\) yields the same power-law behavior, with a normalization consistent with the expected \(M_{\rm PBH}^{3/4}\) scaling.
Both the normalization and the power-law index are in excellent agreement with the analytical induced-halo solution of Ref.\,\cite{Berezinsky:2013fxa}, which describes collisionless dark matter accreting from an initially homogeneous background onto a compact central seed. Our result is also consistent with the PBH-seeded halo prescription of Ref.\,\cite{Ricotti:2007au}, in which the dark-matter envelope grows through secondary accretion with \(M_{\rm UCMH}\propto M_{\rm PBH}(1+z)^{-1}\) and \(R_{\rm UCMH}\propto M_{\rm UCMH}^{1/3}(1+z)^{-1}\). When combined with the standard \(r^{-9/4}\) density profile normalized to the enclosed halo mass, and using \(z_{\rm eq}\simeq3400\), this prescription yields the same \(M_{\rm PBH}^{3/4}r^{-9/4}\) dependence and a normalization in good agreement with our numerical reconstruction.

These agreements show that, for the benchmark scenario considered here, the QCD-axion isocurvature rms fluctuations and the initial departure of the shells from pure Hubble expansion do not produce a discernible modification of the final density profile. Thus, after explicitly incorporating these QCD-axion-specific initial conditions, the minihalo evolution remains governed predominantly by the gravitational seed provided by the central PBH and reproduces the standard secondary-infall result. 

\subsection{Inner density profile and core radius}
\label{Sec:Innercore}

The power-law density profile given in Eq.\,\eqref{eq:rho94num},  $\rho_{\rm halo}(r) \propto r^{-9/4}$,  cannot be extrapolated indefinitely toward the central PBH. At sufficiently small radii, angular-momentum conservation leads to a breakdown of the purely radial-infall approximation, as discussed in Ref.~\cite{Bringmann:2011ut}. The structure of the innermost PBH-dominated region is not uniquely established. Conventional analytic UCMH studies often replace the breakdown of radial infall by a constant-density core \cite{Bringmann:2011ut,2012PhRvD..86d3519L}. More detailed phase-space calculations around PBHs instead obtain broken inner cusps with asymptotic behaviors such as 
$\rho_{\rm halo}(r) \propto r^{-3/4}$ and $\rho_{\rm halo}(r) \propto r^{-3/2}$ whose occurrence and transition radii depend on the PBH mass, the particular dark-matter particle properties, and its kinetic history \,\cite{Boudaud:2021irr}. 

From a phenomenological point of view, it is therefore well motivated to parameterize the unresolved inner axion-minihalo profile as
\begin{equation}
\rho_{\rm halo}(r)=
\rho_{\rm core}
\begin{cases}
\left(\dfrac{r}{r_{\rm core}}\right)^{-\alpha},
& r_\mrm{ISCO}\leq r<r_{\rm core},
\\[2mm]
\left(\dfrac{r}{r_{\rm core}}\right)^{-9/4},
& r\geq r_{\rm core},
\end{cases}
\label{eq:rho_halo}
\end{equation}
where $r_\mrm{ISCO} =6GM_{\rm PBH}$ is the radius of the innermost stable circular orbit,
$0\leq\alpha<9/4$, $\rho_{\rm core} \equiv \rho_{\rm halo} (r_{\rm core})$ from Eq.\,\eqref{eq:rho94num}, and $r_{\rm core}$ is the characteristic radius  at which angular momentum invalidates the purely radial-infall approximation. The density in the innermost region of the minihalo is particularly relevant for the multimessenger signal considered here. A larger local density enhances the dynamical-friction force acting on the neutron star and therefore increases the gravitational-wave dephasing. It also increases the axion flux incident on the neutron-star magnetosphere and hence the resulting radio emission from axion--photon conversion. We adopt the limiting case $\alpha=0$,
corresponding to a constant-density core, as our fiducial and minimally concentrated inner profile. At fixed matching density \(\rho_{\rm halo}(r_{\rm core})\), any choice \(\alpha>0\) increases the density at \(r<r_{\rm core}\) and would therefore generally strengthen the multimessenger signatures considered below.
A self-consistent calculation for \(\alpha>0\) would require revisiting both the density and velocity distributions in the inner halo. Similar constructions to the profile given in \cref{eq:rho_halo} have been used to model the spiky DM profiles surrounding astrophysical black holes~\cite{Eda:2013gg,Eda:2014kra,Edwards:2019tzf}.

We now estimate the characteristic radius \(r_{\rm core}\). The conventional PBH-seeded halo mass and radius prescriptions, $M_\text{UCMH}$ and $R_\text{UCMH}$, are taken from Ref.\,\cite{Ricotti:2007au}, while the velocity-dispersion prescription employed below follows Refs.\,\cite{Bringmann:2011ut,Ricotti:2009bs, Ricotti:2007au}.
The conventional PBH-seeded halo mass and radius should not be identified exactly with the shell-defined minihalo mass and virial radius introduced above. Nevertheless, when combined with the standard \(r^{-9/4}\) profile normalized to the enclosed halo mass, the conventional PBH-seeded prescription reproduces both the mass scaling and, to good accuracy, the normalization of our numerical result, Eq.\,\eqref{eq:rho94num}. We therefore use the standard quantities $M_{\rm UCMH}(z)$, $R_{\rm UCMH}(z)$, and $\sigma_\text{DM}(z)$ consistently in the angular-momentum estimate.

During matter domination, the conventional UCMH mass-growth and radius relations are
\begin{align}
M_{\rm UCMH}(z) &=\frac{1+z_\text{eq}}{1+z} M_{\rm PBH}\,,\label{eq:Mucmhz}\\
R_{\rm UCMH}(z) &= 0.019\,\text{pc} \left( \frac{1000}{1+z} \right)\left(\frac{M_{\rm UCMH}(z)}{M_{\odot}} \right)^{1/3}\,,\label{eq:Rucmhz}
\end{align}
where $z_\text{eq}\simeq 3400$ is the redshift at matter-radiation equality.
Following the standard non-radial-infall treatment, we associate the infalling dark matter with the characteristic specific angular momentum
\begin{equation}
j(z_\sigma) \simeq \sigma_{\text{DM}}(z_\sigma)R_{\text{UCMH}}(z_\sigma)\,,
\end{equation}
where \(\sigma_{\rm DM}\) is the transverse velocity dispersion and $z_{\sigma} \sim 1000$ is the redshift at which the UCMH inner most region collapses~\cite{Ricotti:2009bs}. Conservation of angular momentum then gives the characteristic tangential velocity
\begin{equation}
v_{\text{rot}}(r,z_\sigma) = \frac{j(z_\sigma)}{r} = \frac{\sigma_{\text{DM}}(z_\sigma)R_{\text{UCMH}}(z_\sigma)}{r}\,.\\
\label{eq:vrot}
\end{equation}
The conventional UCMH calculation compares this velocity with the circular velocity generated only by the enclosed halo mass. In the innermost region of a PBH-seeded halo, however, the central PBH must be included. The appropriate circular velocity is
\begin{equation}
v^2_{\text{cir}}(r,z_\sigma) = \frac{G(M_\text{PBH}+M_\text{halo}(<r,z_\sigma))}{r}\,.
\label{eq:vcir}
\end{equation}
The non-radial-infall scale is defined by the condition
$v_{\text{rot}}(r_\text{core},z_\sigma) = v_{\text{cir}}(r_\text{core},z_\sigma)$. Since the solution lies deeply inside the PBH-dominated region, we take
  $M_\text{PBH}+M_\text{UCMH}(r<r_\text{core},z_\sigma)\simeq M_\text{PBH}$. Thus,
\begin{equation}
r_\text{core} \simeq \frac{\sigma^2_\text{DM}(z_\sigma)R^2_\text{UCMH}(z_\sigma)}{G M_\text{PBH}}\,.
\label{eq:rcore1}
\end{equation}
Using the analytic quantities \(M_{\rm UCMH}(z)\) and \(R_{\rm UCMH}(z)\) as an effective parameterization of the outer halo, Eqs.\,\eqref{eq:Mucmhz} and \eqref{eq:Rucmhz}, and the characteristic DM velocity dispersion from Refs.\,\cite{Ricotti:2009bs, Ricotti:2007au},
\begin{equation}
\sigma_\text{DM}(z) = 0.14\,\text{m/s} \left(  \frac{1000}{1+z}\right)^{1/2}\left(  \frac{M_\text{UCMH}(z)}{M_\odot}\right)^{0.28}\,,
\end{equation}
we obtain
\begin{align}
r_{\rm core} \simeq {}&
7\times10^{-9}\,{\rm pc}
\left(\frac{1+z_\sigma}{1000}\right)^{-4.227}
\nonumber\\
&\times
\left(\frac{1+z_{\rm eq}}{3400}\right)^{1.227}
\left(\frac{M_{\rm PBH}}{M_\odot}\right)^{0.227}.
\label{eq:rcore}
\end{align}
For the fiducial values $z_\sigma=1000$ and $z_\text{eq}=3400$, we have $r_\text{core} \simeq (7\times10^{-9} -  1 \times 10^{-8})\,\text{pc}$ with corresponding core densities $\rho_\mrm{core} \simeq (2\times10^{-4}\text{--}6\times10^{-4})\,\mrm{g\,cm^{-3}}$ for $M_{\text{PBH}}=(1,10)\,M_\odot$, respectively.

\section{Multi-messenger signal during neutron star-axion/UCMH encounters}

We explore the multimessenger emission generated during an encounter between a neutron star and a QCD-axion ultracompact minihalo, with particular emphasis on the inspiral of the neutron star around the central PBH. Throughout this phase, the system produces an appreciable gravitational-wave signal, which can be modified by the presence of the QCD-axion minihalo. Simultaneously, QCD axions entering the NS magnetosphere may resonantly convert into photons in the strong ambient magnetic field. These two messengers together define a distinctive observational signature, carrying information about the characteristic properties of both the UCMH and the NS. The gravitational wave signal can be used to infer the presence of the UCMH around the PBH from the dephasing of the GW template compared to the bare vacuum case. The electromagnetic signal instead probes the QCD-axion particle component of the minihalo through narrow-band radio emission, with a characteristic peak frequency determined by the axion mass.

\subsection{Gravitational wave signature}
\label{sec:gw_sig}

\begin{figure}
    \centering
    \includegraphics[width=0.48\textwidth]{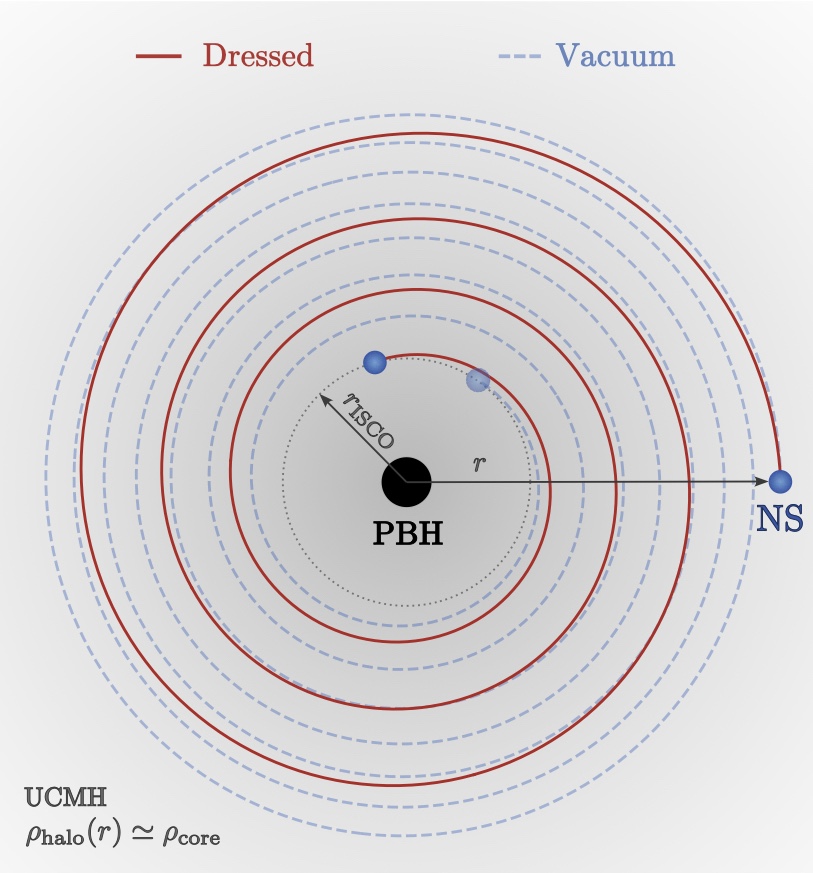}
    \caption{Schematic illustration of how an ultracompact minihalo seeded by a primordial black hole impacts the inspiral dynamics of a PBH-NS merger. The dressed PBH induces an additional energy loss mechanism due to dynamical friction, leading to the accelerated infall of the NS and thus a dephasing of the gravitational wave signal compared to the vacuum case. The trajectories are shown down to the ISCO radius $r_\mrm{ISCO}$. For the portion of the inspiral relevant for our work, the density of the UCMH is approximately constant at the core density, $\rho_\mrm{halo}(r) \simeq \rho_\mrm{core}$.}
    \label{fig:gw_diagram}
\end{figure}

The presence of an axion minihalo around a central
PBH modifies the inspiral dynamics via two distinct
mechanisms. Firstly, as the neutron star orbits the PBH, it experiences dynamical friction due to its movement through the minihalo. This provides an additional energy-loss channel beyond GW emission, accelerating the orbital decay. Secondly, the additional enclosed mass provided by the minihalo leads to a deviation from the point-like Keplerian orbit predicted by the sole presence of the central PBH, leading to a frequency shift of the GW. Both of these effects result in a dephasing of the GW waveform compared to the vacuum case; however, the second of these is negligible compared to the first in our case.\footnote{For instance, the DM mass contained within the core radius is $M_\mrm{halo} = 4/3 \pi r_\mrm{core}^3 \rho_\mrm{core}$. Taking the $M_\mrm{PBH} = 1\,M_\odot$ benchmark, we find that $M_\mrm{halo} / M_\mrm{PBH} \sim 10^{-5}\ll1$. We therefore neglect the mass of the halo when computing the dephasing, which remains negligible in the regime we analyze here.} This dephasing is our ultimate signature for the presence of the axion minihalo. 

We illustrate the effect of the UCMH on the PBH-NS inspiral in \cref{fig:gw_diagram}. The UCMH dressing the PBH provides an additional energy loss mechanism to the infalling NS through dynamical friction, resulting in a faster merger compared to the vacuum case. This leads to a dephasing of the GW waveform that can lead to a significant loss of sensitivity if the incorrect template is used when reconstructing the signal; this can be used as a smoking gun signature for the presence of the UCMH. This mismatch relies on the integrated phase difference between the templates over many orbital cycles. In the figure, we exaggerate the effect for visual clarity, as the dephasing we have sensitivity to can be of the order of only one radian.

The GW phase $\varphi(t)$ enters in the waveform $h(t) = h_0 \cos[\varphi(t)]$. This phase accumulates as
\begin{equation}
    \varphi = 2\pi\int_{t_{0}}^{t} f(t') \dd t' = 2\pi \int_{f_\mrm{min}}^f \frac{f'}{\dot{f}'    } \dd f'\,,
\label{eq:gw_phase}
\end{equation}
where $t_0$ is the initial time of observation, corresponding to a minimum observed GW frequency of $f_\mrm{min}$, when the inspiraling bodies are farthest away. For approximately circular orbits in the Newtonian regime, the GW frequency is given by twice the orbital frequency, such that 
\begin{equation}
\label{eq:gw_frequency}
    f = 2
    f_\mrm{orb} = \frac{1}{\pi} \sqrt{\frac{G M_\mrm{tot}}{r^3}}\implies \frac{f}{\dot{f}} = - \frac{2}{3} \frac{r}{\dot{r}}\,,
\end{equation}
where $M_\mrm{tot} \equiv M_\mrm{PBH} + M_\mrm{NS}$ is the total mass of the binary system and $r$ is the distance between them. Solving for the accumulated phase therefore amounts to solving the equation of motion for the infalling binary system. 

We solve this equation of motion by following a similar treatment to Refs.~\cite{Eda:2014kra,Edwards:2019tzf}, considering the loss of energy in the system due to both dynamical friction and GW emission. As we will show below, the portion of the inspiral relevant for the most sensitive LISA frequency range occurs inside the UCMH core, \(r<r_{\rm core}\). Hence, for our fiducial flat-core profile, the ambient DM density entering the dynamical-friction calculation is simply
$\rho_{\rm halo}(r<r_{\rm core})=\rho_{\rm core}$.

The effect of dynamical friction can be treated classically via the Chandrasekhar formula as long as the axion field behaves as a collisionless fluid as the neutron star passes through it~\cite{Chandrasekhar:1943ys}. We establish this condition by requiring that the deBroglie wavelength of the axion field $\lambda_\mathrm{dB}$ be much smaller than the characteristic gravitational radius of the neutron star $R_g$. We then have that
\begin{equation}
\begin{split}
\frac{\lambda_\mathrm{dB}}{R_g} &= \left(\frac{2\pi}{m_a v_\mrm{NS}}\right) \left(\frac{v_\mrm{NS}^2}{G M_\mathrm{NS}}\right)\\
&\sim 10^{-4}\left(\frac{10^{-6} \uni{eV}}{m_a}\right)\left(\frac{1.4\,M_\odot}{M_\mrm{NS}}\right)\left(\frac{v_\mrm{NS}}{0.4}\right) \ll 1\,,
\end{split}
\end{equation}
where $v_\mrm{NS}$ is the orbital velocity of the NS, which also acts as the approximate velocity of the axion field with respect to the NS. In the above, we have fiducialized to the velocity at the innermost stable circular orbit and the typical mass of an NS.  This shows that we are well within the classical limit.

In the classical regime, the dynamical friction caused by the presence of the UCMH arises from a dissipative force that acts in the opposite direction to the NS velocity. It may be written as~\cite{Chandrasekhar:1943ys}
\begin{equation}
    F_\mrm{DF}(r) = \frac{4 \pi G^2 M_\mrm{NS}^2 \rho_{\rm core}}{v_\mrm{NS}^2(r)} \ln \Lambda\,,
\end{equation}
where $\ln \Lambda$ is the Coulomb logarithm, which we take to be $\ln \Lambda = 3$~\cite{Edwards:2019tzf}. This force leads to an additional energy loss beyond the usual GW emission, given by 
\begin{equation}
    \dot{E}_{\mathrm{DF}} = F_\mrm{DF} v_\mrm{NS} =\frac{4 \pi G^{3 / 2} M_\mrm{NS}^2 M_{\mathrm{tot}}^{1 / 2} \rho_{\rm core} \ln \Lambda}{M_{\mathrm{PBH}}} r^{1/2} \,,
\end{equation}
where $v_\mrm{NS}(r) = r_\mrm{NS} \omega_\mrm{orb}$ is the orbital velocity of the NS relative to the barycenter of the system, with $r_\mrm{NS} = (M_\mrm{PBH} / M_\mrm{tot}) r$ the distance to the NS from the barycenter of the system and $\omega_\mrm{orb} = (G M_\mrm{tot} / r^3)^{1/2}$ the orbital angular velocity. 
Alongside this energy loss is that from the usual emission of GWs. For circular orbits in the Newtonian regime, this is given by the Peters formula~\cite{Peters:1963ux,Peters:1964zz},
\begin{equation}
    \dot{E}_\mrm{GW} = \frac{32}{5} \frac{G  \mu^2}{c^5}r^4 \omega_\mrm{orb}^6\,.
\end{equation}

The total change in the orbital  energy of the system is given by the energy balance equation $\dot{E}_\mrm{orbit} = -(\dot{E}_\mrm{GW} + \dot{E}_\mrm{DF})$, where the virial theorem readily gives us that
\begin{equation}
      \dot{E}_\mrm{orbit} = \frac{1}{2}\frac{G M_\mrm{tot}\mu}{r^2}\dot{r}\,.
\end{equation}
The energy balance equation can then be re-expressed as
\begin{equation}
    \dot{r} = -\left(\mcal{C}_\mrm{GW} r^{-3} + \mcal{C}_\mrm{DF}r^{5/2}\right)\,,
    \label{eq:r_dot}
\end{equation}
where we have defined the GW and dynamical friction coefficients
\begin{equation}
\begin{split}
    \mcal{C}_\mrm{GW} &\equiv \frac{64}{5} \frac{G^3  M_{\mathrm{tot}}^2\mu}{c^5}\,,\\
    \mcal{C}_\mrm{DF} &\equiv \frac{8 \pi G^{1 / 2} M_\mrm{NS}^2  \rho_{\rm core} \ln \Lambda}{M_{\mathrm{PBH}}M_{\mathrm{tot}}^{1 / 2}\mu}\,.
 \label{eq:CGW}   
\end{split}
\end{equation}
The GW term dominates for orbital radii less than the crossover radius
\begin{equation}
  r \ll r_{*} \equiv
  \left( \frac{\mathcal{C}_{\mathrm{GW}}}{\mathcal{C_{\mathrm{DF}}}} \right)^{2/11}\,,
  \label{eq:rco}
\end{equation}
which takes the values $r_{*} \approx (2.8 
\times 10^{-10}-1.0 \times 10^{-9})\uni{pc}$ for
$M_\text{NS}=1.4\,M_\odot$ and
$M_\text{PBH}=(1,10)\,M_\odot$, respectively,  where $r_{*}/r_\text{core} \sim 10^{-2}$. For the binary system under consideration, the range of GW frequencies associated with the UCMH core correspond to  $1.8 \uni{mHz} \lesssim f \lesssim 1.8 \uni{kHz}$ and $2.3 \uni{mHz} \lesssim f \lesssim 0.39 \uni{kHz}$ for $M_\mrm{PBH}=(1,10)\, M_\odot$, respectively. Here, the lower and upper frequencies in each case are \(f(r_{\rm core})\) and \(f_{\rm ISCO}\), the gravitational-wave frequency at the innermost stable circular orbit~\cite{Maggiore:2007ulw}. 

Inserting our expression for $\dot{r}$ in \cref{eq:r_dot} into \cref{eq:gw_phase}, we find that the phase evolves as
\begin{equation}
    \varphi(f) = 2\pi \int_{f_\mrm{min}}^f\frac{2 r^4(f')}{3 \mcal{C}_\mrm{GW}} \left[1 + \left(\frac{r}{r_{*}}\right)^{11/2}\right]^{-1} \dd f'\,,
   \label{eq:phaseevolves} 
\end{equation}
We identify the dephasing by the DM halo as $\Delta \varphi \equiv \varphi- \varphi_\mrm{vac}$, where the purely GW term, $\varphi_\mrm{vac}$, is obtained from 
$\dot r =  -\mathcal{C}_\text{GW}r^{-3}$. Thus,
\begin{align}
    \Delta\varphi(f) &= \frac{4(G M_\text{tot})^{4/3}}{5\pi^{5/3}\mathcal{C}_\text{GW}}\\
    &\times
    \left[ f_\text{min}^{-5/3}(\mathcal{H}(f_\text{min})-1)-f^{-5/3}(\mathcal{H}(f)-1)  \right]\,, \nonumber
\end{align}
where
\begin{equation}
    \mathcal{H}(f)\equiv \textcolor{white}{i}_2F_1\left( 1, \frac{5}{11};\frac{16}{11}, -\frac{(GM_\text{tot})^{11/6}}{\pi^{11/3}r_{*}^{11/2}} f^{-11/3}\right)
\end{equation}
is the hypergeometric function.

\begin{figure}
    \centering
    \includegraphics[width=0.48\textwidth]{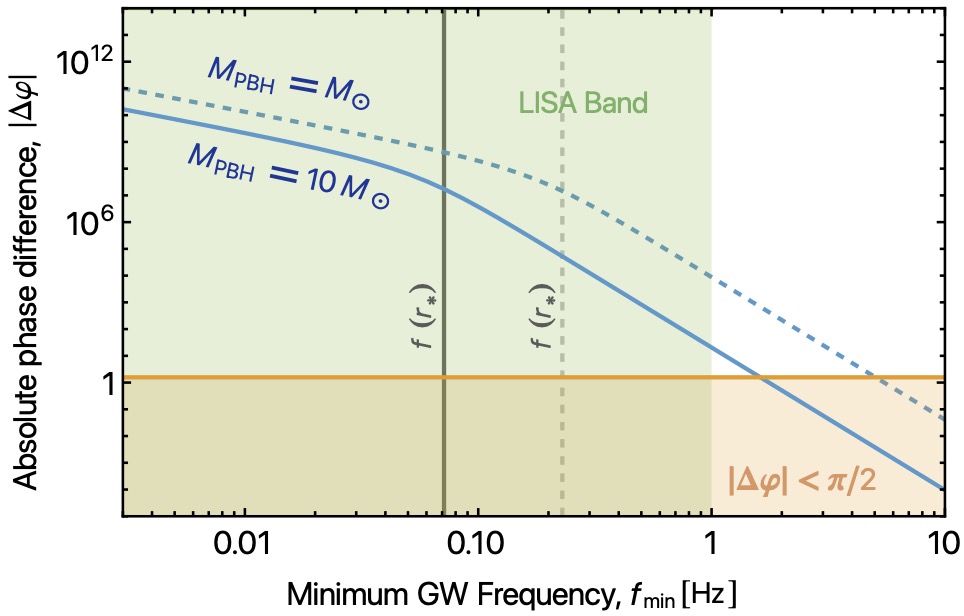}
    \caption{The total absolute phase difference $|\Delta \varphi|$ in the gravitational wave signature from a PBH-NS merger in the presence of an axion minihalo compared to the vacuum scenario. This is shown with respect to the smallest gravitational wave frequency $f_\mrm{min}$ observed by an experiment during the inspiral phase. The results are computed for PBHs with masses $1\,M_\odot$ (dashed blue line) and $10\,M_\odot$ (solid blue line), and an NS of mass $M_\mrm{NS} = 1.4\,M_\odot$. The green region shows part of LISA frequency band~\cite{LISA2017,LISA:2024hlh}. The orange region highlights where $|\Delta \varphi| <\pi/2$, where we do not expect to be able to detect the presence of the minihalo. The two vertical lines refer to the turnover frequencies $f(r_{*})$ at which the energy loss from either gravitational wave emission or dynamical friction dominate for primordial black holes with masses $1\,M_\odot$ (dashed black line) and $10\,M_\odot$ (solid black line). }
    \label{fig:phase_gw}
\end{figure}

We show the total absolute accumulated phase difference $\Delta \varphi$ as a function of the minimum observed GW frequency $f_\mrm{min}$ in \cref{fig:phase_gw}, taking the final frequency to be $f = f_\mrm{ISCO} \gg f_\mrm{min}$. We illustrate this for an NS-PBH binary system with an NS of mass $M_\mrm{NS} = 1.4\,M_\odot$ and a PBH of mass $M_\mrm{PBH} = 1\,M_\odot$ or $M_\mrm{PBH} = 10\,M_\odot$, embedded in the core of a QCD axion UCMH. As we show in Appendix\,\ref{app:matched_gw}, a total accumulated phase in excess of $\Delta \varphi \gtrsim \pi/2$ in the GW signal results in a reconstructed signal-to-noise ratio that is at least halved when using the incorrect vacuum signal template in a matched filter analysis. We take this phase as the detectability threshold for a UCMH, as it leads to a significant loss in the reconstructed sensitivity.

Since the dephasing of the GW waveform compared to the vacuum expectation is greater at larger radii, where dynamical friction plays an increasingly important role, the effect is more pronounced at lower GW frequencies. This motivates us to search for this effect using the planned low-frequency GW detector LISA, which will search for GWs in the frequency range $f = 0.1\,\mrm{mHz}\text{--}1\,\mrm{Hz}$~\cite{LISA:2024hlh,LISA2017}. For PBHs in the mass range considered here, \(M_{\rm PBH}=(1\text{--}10)M_\odot\), a significant dephasing extends over the LISA frequency range. 

\cref{fig:phase_gw} shows that there are two regimes for the GW dephasing signature, corresponding to $r \gtrsim r_{*}$ and $r \lesssim r_{*}$. At larger radii (lower frequencies), dynamical friction dominates the energy loss of the binary system, whereas at smaller radii (higher frequencies) GW emission becomes the dominant energy-loss mechanism. Since the GW dephasing is sourced by dynamical friction, the accumulated dephasing is much larger when the observation begins in the DF-dominated regime, \(f_{\min}\ll f_{*}\), than when it begins deep in the GW-dominated regime, \(f_{\min}\gg f_{*}\), where $f_{*} \equiv f(r_{*})$. For sufficiently small radii, the GW dephasing falls below the $\pi/2$ threshold and approaches zero. For PBH masses much larger than $10\,M_\odot$, this means that the dephasing signature eventually becomes entirely hidden to LISA.

These two regimes have an asymptotic characteristic slope, which measures the rate at which the GW dephasing occurs with the minimum measured frequency. Taking the differential form of \cref{eq:phaseevolves} and defining the parameter $\eta \equiv (r/r_{*})^{11/2}= (f_{*}/f)^{11/3}$, we have that
\begin{equation}
\frac{\dd \varphi}{\dd f} = \frac{\mathcal{K}f^{-8/3} }{1+\eta}\,,
\end{equation}
where $\mathcal{K} \equiv 4(G M_\text{tot})^{4/3}/(3\pi^{5/3}\mathcal{C}_\text{GW})$  and  $\dd\varphi_\text{vac}/\dd f$ is retrieved by setting $\eta = 0$. The corresponding differential phase difference is then
\begin{equation}
\frac{\dd(\Delta\varphi)}{\dd f} = -\mathcal{K}f^{-8/3}\frac{ \eta}{1+\eta}\,.
\end{equation}
For high frequencies $f \gg f_{*}$, we have $\eta \ll 1$. Thus
$\dd (\Delta\varphi)/\dd f \propto -f^{-8/3} \eta \propto f^{-19/3}$. Integrating from
$f_\text{min}$ to $f_\text{ISCO}$, with $f_\text{ISCO} \gg f_\text{min}$, we then have $|\Delta  \varphi|\propto f^{-16/3}_\text{min}$. This  corresponds to the steep branch on the right side of each curve. Even though the binary energy loss driven by dynamical friction is subleading, its fractional importance relative to GW emission changes quickly since 
$\eta \propto f^{-11/3}$. 

For low frequencies, $f \ll f_{*}$, we have $\eta \gg 1$
so that $\dd \Delta\varphi/\dd f \simeq -\mathcal{K}f^{-8/3}=-\dd \varphi_\text{vac}/\dd f$. After integration, $|\Delta  \varphi|\propto f^{-5/3}_\text{min}$. This corresponds to the shallower low-frequency branches in \cref{fig:phase_gw}. When the dynamical friction becomes dominant, the binary evolves much faster than in vacuum and accumulates comparatively little phase at low frequencies. The dephasing then becomes essentially set by the vacuum phase that would otherwise have accumulated over the same frequency interval, leading to the scaling \(|\Delta\phi| \propto f_{\min}^{-5/3}\).

In summary, the asymptotic behavior of the gravitational dephasing regarding the dynamical friction and gravitational wave  processes obey
the functional form
\begin{equation}
|\Delta\phi| \propto
\begin{cases}
f_{\min}^{-5/3}
&\text{if } f_{\min}\ll f_{*}
\\
& \hspace{-1.5cm}\text{DF dominated}\,,
\\[1mm]
f_{\min}^{-16/3},
&\text{if } f_{\min}\gg f_{*}
\\
& \hspace{-1.5cm}\text{GW dominated, perturbative DF}\,.
\end{cases}
\end{equation}

Ref.~\cite{Eda:2014kra, Edwards:2019tzf} explored the GW signature from DM minihalos accounting for both dynamical friction and deviations from Keplerian orbital evolution. Their analysis focused on a DM spike seeded by an intermediate-mass black hole of mass $\usim 10^3\,M_\odot$. This produced power-law density profiles within spike radii of $r_\mathrm{sp} \sim 1\,\mathrm{pc}$ and with  characteristic normalizations $\rho_\mathrm{sp} \sim 10^2\,M_\odot\,\mathrm{pc}^{-3}$, following an NFW profile beyond the spike. The UCMHs we consider here are qualitatively different. As discussed in the Introduction, the outer \(r^{-9/4}\) envelope of PBH-seeded minihalos is a robust prediction of the secondary-infall scenario\,\cite{1985ApJS...58...39B}. It has been obtained analytically for generic collisionless dark matter\,\cite{Berezinsky:2013fxa, Ricotti:2007au}, reproduced in numerical simulations\,\cite{Adamek:2019gns}, and recovered here in our explicit calculation for QCD-axion dark matter. By contrast, the structure of the angular-momentum-supported inner region is less certain, with no unique prediction for its density profile (see discussion in Sec.\,\ref{Sec:Innercore}). We therefore adopt a minimally concentrated, constant-density core as our fiducial benchmark. Even in this conservative case, the matching to the steep outer envelope yields a very large core density, \(\rho_{\rm core}\sim10^{19}\,M_\odot\,{\rm pc}^{-3}\), which produces a detectable GW dephasing for the benchmark systems we consider. In particular, although a flat core with the much lower normalizations considered in Ref.~\cite{Eda:2014kra} would lead to an undetectable effect, the extreme densities predicted for PBH-seeded UCMHs place the corresponding dephasing within the reach of future detectors.

\subsection{Radio signatures from resonant axion--photon conversion}

The radio counterpart of our multi-messenger framework is produced by the resonant conversion of axions from the UCMH as they enter the neutron-star magnetosphere. This process generates a narrow spectral line whose flux and time-dependent broadening encode the local axion density, the axion–photon coupling, and the binary orbital motion. Consequently, the radio signal provides a direct probe of the particle nature of the minihalo, complementary to the gravitational-wave dephasing.

\subsubsection{Neutron Star Model}

We represent the neutron star as a spherical body of radius \(R_{\rm NS}\), rotating with angular velocity \(\boldsymbol{\Omega} = \Omega {\hat{\bv{z}}}\) about an axis through its center. The magnitude of this  spin is $\Omega = 2\pi / P$,
where \(P\) is the rotational period. In the co-rotating frame, the magnetosphere is approximated by a dipolar magnetic field generated by a magnetic moment \(\mathbf{m}(t)\) of constant magnitude \(m\) and a potentially time-dependent direction. The magnetic field is therefore
\begin{equation}
\mathbf{B}(\mathbf{r},t)=\frac{B_0\big[3(\hat{\mathbf{m}}(t)\!\cdot\!\hat{\mathbf{r}})\hat{\mathbf{r}} - \hat{\mathbf{m}}(t)\big] }{2}
\left(\frac{r}{R_{\rm NS}}\right)^{-3},
\label{eq:Bgeneral}
\end{equation}
where $B_0 \equiv 2m/R_{\rm NS}^3$ is the magnetic
field amplitude at the surface poles.

In the Goldreich--Julian framework\,\cite{1969ApJ...157..869G}\,\footnote{This framework yields the lowest plasma density required for charges moving along dipolar magnetic field lines to remain in co-rotation with the star. The Goldreich–Julian description has been widely employed in earlier work on axion dark matter \,\cite{Millar:2021gzs, Foster:2020pgt, Leroy:2019ghm, Safdi:2018oeu, Huang:2018lxq, Pshirkov:2007st}, as well as in studies of solitons\,\cite{Schiappacasse:2026ems, Amin:2021tnq, Witte:2022cjj, Buckley:2020fmh}, miniclusters\,\cite{Witte:2022cjj,Edwards:2020afl}, minihalos surrounding primordial black holes\,\cite{Choi:2022btl, Nurmi:2021xds}, and axion dark radiation\,\cite{Long:2024qvd}.}, the local charge density determines the number density of magnetospheric charges according to
\begin{equation}
n_c(\mathbf{r},t)=\frac{2\,\boldsymbol{\Omega}\!\cdot\!\mathbf{B}(\mathbf{r},t)}{e},
\end{equation}
where \(e\simeq 0.303\). Outside the polar regions, where a non-relativistic electron--proton plasma provides a reasonable approximation, the plasma frequency is controlled by the electron density,
\begin{equation}
\omega_{\rm pl}^2(\mathbf{r},t)=\frac{e^2\,n_e(\mathbf{r},t)}{m_e},
\end{equation}
with \(m_e\simeq 0.511\,{\rm MeV}\) the electron mass.
We adopt the minimal charge-separated Goldreich–Julian prescription, assuming that the local corotation charge density $n_c$ is supplied predominantly by a single light charge species, so that \(n_e\simeq |n_c|\)\,\cite{Long:2024qvd, Hook:2018iia}. Therefore, the plasma frequency takes the form
\begin{equation}
\omega_{\rm pl}(\mathbf{r},t)
=
\omega_{{\rm pl},0}  \,
\Bigg\lvert\frac{2\boldsymbol{\hat\Omega} \cdot \mathbf{B}(\mathbf{r}, t)}{B_0}\Bigg\lvert^{1/2}\,,
\label{eq:wpl}
\end{equation}
where $\omega_{{\rm pl},0} \equiv (e \Omega B_0/m_e)^{1/2}$ is the surface plasma frequency at the magnetic equator in the aligned configuration. For characteristic galactic neutron-star parameters \,\cite{Safdi:2018oeu,Faucher-Giguere:2005dxp}, \(B_0\sim10^{13}\,{\rm G}\) and \(P\sim3\,{\rm s}\), this normalization is of order \(\omega_{\rm pl,0}\sim 10\,\mu{\rm eV}\).

\begin{figure}[t]
\centering
 \includegraphics[width= 70mm]{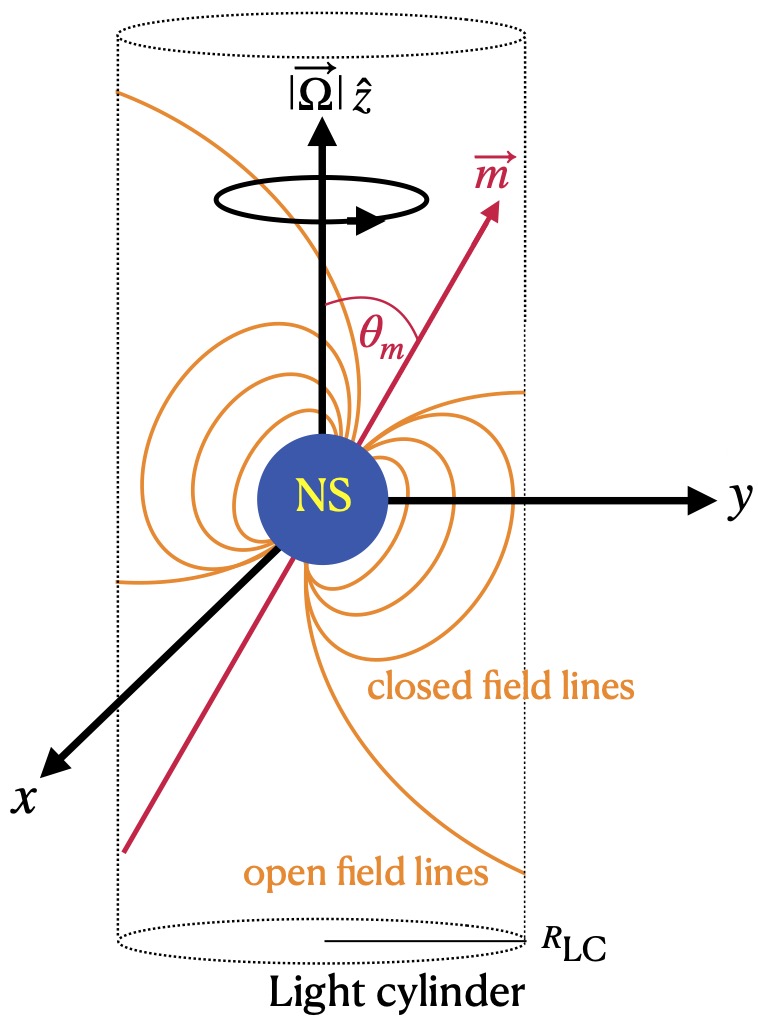}
\caption{Schematic illustration for the magnetic field lines around a rotating neutron star with angular velocity $\boldsymbol{\Omega} = \Omega \,\hat{\mathbf{z}}$, where $\hat{\boldsymbol{\Omega}}\cdot\hat{\mathbf{m}} = \text{cos}(\theta_m)$. This figure has been adapted from Ref.\,\cite{Schiappacasse:2026ems}  with the permission of the authors. }
 \label{fig:NSdiagram}
\end{figure}

The above description is valid only within the light cylinder
\begin{equation}
R_{\rm LC}\simeq 4.8\times10^4\,{\rm km}
\left(\frac{P}{1\,{\rm s}}\right)
\end{equation}
and is expected to hold in the radial interval $R_{\rm NS}<r\ll R_{\rm LC}$
 as long as one stays away from the polar caps, where relativistic plasma effects can become important\,\cite{Safdi:2018oeu, Long:2024qvd}. \cref{fig:NSdiagram} presents a schematic illustration of a neutron star together with its magnetosphere, with the rotation axis aligned along the 
$z$-axis and the magnetic dipole inclined by an angle 
$\theta_m$
 relative to it.

\begin{table*}[t]
\centering
\caption{Neutron-star benchmarks used in our analysis. Daggers denote model-dependent or fiducial quantities rather than direct observables. The M31 source is an accreting X-ray pulsar and is used only as an extragalactic-distance benchmark. Here RQ denotes radio quiet, Fid. denotes a fiducial population benchmark, and XRB denotes an accreting X-ray binary.}
\label{tab:NStargets}
\resizebox{\textwidth}{!}{%
\begin{tabular}{lccccccc}
\hline\hline
Source & Status & \(d\) [pc] & \(P\) [s] & \(B_0\) [G] 
& \(M_\text{NS}\,[M_\odot]\) & \(R_\text{NS}\) [km] & \(\theta_m\) \\
\hline
RX~J1856.5$-$3754 
& RQ
& \(123^{+11}_{-15}\) \cite{Walter:2010}
& \(7.055\) \cite{vanKerkwijk:2007}
& \(1.5\times10^{13\,\dagger}\) \cite{vanKerkwijk:2007}
& \(1.48^{\dagger}\) \cite{Potekhin:2014}
& \(12.1^{\dagger}\) \cite{Potekhin:2014}
& \(\lesssim6^\circ{}^{\dagger}\) \cite{Potekhin:2014} \\

RX~J0806.4$-$4123 
& RQ
& \(250^{\dagger}\) \cite{Posselt:2007}
& \(11.37\) \cite{Kaplan:2009}
& \(2.5\times10^{13\,\dagger}\) \cite{Kaplan:2009}
& \(1.4^{\dagger}\) \cite{Haberl:2002vx}
& \(10^{\dagger}\) \cite{Haberl:2002vx}
& \(15^\circ{}^{\dagger}\) \cite{Hook:2018iia} \\

RX~J0720.4$-$3125 
& RQ
& \(360^{+170}_{-90}\) \cite{Kaplan:2007}
& \(8.39\) \cite{Zane:2002}
& \(2.1\times10^{13\,\dagger}\) \cite{Zane:2002}
& \(1.30^{\dagger}\) \cite{Luo:2022itt}
& \(12.3^{\dagger}\) \cite{Luo:2022itt}
& \(30^\circ\!\,^{\dagger}\) \cite{Perez-Azorin:2006htq} \\

Typical galactic NS
& Fid.
& \(10^3{}^{\dagger}\)
& \(3^{\dagger}\) \cite{Safdi:2018oeu,Faucher-Giguere:2005dxp}
& \(10^{13\,\dagger}\) \cite{Safdi:2018oeu,Faucher-Giguere:2005dxp}
& \(1.4^{\dagger}\)
& \(10^{\dagger}\)
& \(30^\circ{}^{\dagger}\) \\

3XMM~J004232.1$+$411314
& XRB
& \(7.8\times10^5\) \cite{Holland:1998br}
& \(3.0\) \cite{Castillo:2018fyg}
& \(10^{13\,\dagger}\) \cite{Castillo:2018fyg}
& \(1.5^{\dagger}\)
& \(10^{\dagger}\)
& \(30^\circ{}^{\dagger}\) \\
\hline\hline
\end{tabular}%
}
\end{table*}

The QCD axion-photon resonant conversion occurs in a surface within the neutron star magnetosphere at which $m_a\approx \omega_{\rm pl}(r_c)$. Using \cref{eq:wpl}, we have\,\cite{Long:2024qvd} 
\begin{align}
r_c &\approx 17 R_\text{NS} \left(  \frac{B_0}{10^{14}\,\text{G}}\right)^{1/3}
\left(  \frac{P}{1\,\text{s}}\right)^{-1/3}
\left(  \frac{m_a}{10^{-6}\,\text{eV}}\right)^{-2/3}\nonumber\\
&\times \bigg|\hat{\boldsymbol{\Omega}} \cdot \big[3(\hat{\mathbf{m}}(t)\!\cdot\!\hat{\mathbf{r}})\hat{\mathbf{r}} - \hat{\mathbf{m}}(t)\big]\bigg|^{1/3}\,.
\label{eq:rcfull}
\end{align}
As a baseline description of the galactic neutron-star population, we follow the population-synthesis framework given in Refs.\,\cite{Faucher-Giguere:2005dxp,Safdi:2018oeu}. Faucher-Giguère and Kaspi model the neutron-star birth spin periods via a normal distribution with mean $\langle\text{(P/s)}\rangle = 0.3$ and standard deviation $\sigma_{\text{P}} = 0.15$, and the magnetic fields via a log-normal distribution with mean $\langle\text{log}_{10}(B_0\text{G)}\rangle = 12.65$ and standard deviation $\sigma_{\text{log}_{10}(B_0\text{G})} = 0.55$. Later, Safdi et al.~evolve these birth distributions to obtain the present-day neutron-star population. From the peaks of the full-population distributions shown in Fig.~4 of Ref.\,\cite{Safdi:2018oeu}, we adopt the representative values
$P\simeq3\,{\rm s}$ and  $B_0\simeq10^{13}\,{\rm G}$. These values allow us to estimate the radio signal expected from the bulk of the galactic neutron-star population.

In addition, we consider three representative members of the Magnificent Seven neutrons stars as nearby isolated stellar benchmarks: RX J1856.5-3754, RX J0806.4-4123, and RX J0720.4-3125. We tabulate their properties in \cref{tab:NStargets}. These sources are useful because they are radio quiet, strongly magnetized, and comparatively clean targets for applying a minimal Goldreich–Julian magnetospheric description \,\cite{Haberl:2002vx, Posselt:2024}. We stress, however, that not all parameters entering \cref{tab:NStargets} have the same observational status: the spin periods, distances where available, and absence of detected radio pulsations are directly constrained, whereas the dipole field, radius, mass, and magnetic obliquity are partly model dependent.

Our choice is guided by two simple criteria: proximity to an idealized isolated-neutron-star magnetosphere and diversity in source geometry. Among the three benchmarks, RX J1856.5-3754 provides the cleanest case, owing to its small pulsed fraction, nearly thermal spectrum, and long-term stability\,\cite{Haberl:2007, Popov:2003jf}. RX J0806.4-4123 is used as a second relatively clean benchmark, while RX J0720.4-3125 is retained as a useful comparison case with larger systematic uncertainty because of its more pronounced spectral and timing variability \,\cite{Perez-Azorin:2006htq, DeGrandis:2022, Potekhin:2014, Kaplan:2009}. 

For the case of extragalactic NSs, we focus on the closest major galaxy to the Milky Way, the spiral  Andromeda Galaxy M31 located at a distance of $780\, \text{kpc}$\,\cite{Holland:1998br}. To our knowledge\,\footnote{A possible isolated magnetar origin has been discussed for GRB 070201, but no persistent periodic Soft Gamma Repeater/Anomalous X-ray Pulsar counterpart in M31 was identified\,\cite{Mazets:2007gs}.}, no isolated rotation-powered neutron star has been individually identified with the timing information required to infer a standard magnetic-dipole field. The confirmed NS systems in M31 with measured spin periods are instead accreting X-ray pulsars, such as 3XMM J004232.1+411314, which possesses $P\approx 3\,\text{s}$ X-ray pulsations, holding a possible magnetic field ranging from a few $10^{10}$\,\text{G} to $10^{13}\,\text{G}$ depending on its spin period evolution period\,\cite{Castillo:2018fyg}. We therefore use the observed distance and spin period of this known M31 accreting X-ray pulsars only as motivation for an extragalactic neutron-star benchmark, while treating the magnetic field and Goldreich-Julian plasma profile as idealized inputs. 
The numerical values adopted for all the sources mentioned above are summarized in \cref{tab:NStargets}.

\subsubsection{Enhanced Dark Matter density at the conversion radius}
\label{subsec:DM density}

One important ingredient in the calculation of the radio signal emitted through axion–photon conversion during the NS–PBH inspiral is the axion density at the resonant conversion surface. In the previous sections, we reconstructed the UCMH density profile, adopted a fiducial and minimally concentrated inner core for \(r\lesssim r_{\rm core}\), and determined the conversion radius \(r_c\) from the neutron-star magnetospheric plasma profile. However, the density entering the conversion power is not simply the unperturbed minihalo density at the NS orbital position: the neutron-star gravitational potential focuses the incoming axion trajectories and enhances their density near \(r_c\). In this subsection, we estimate this enhanced density by propagating the inner-core phase-space distribution into the neutron-star potential using Liouville’s theorem and then averaging the resulting density over the conversion surface. This quantity provides the local axion-density input required for the radio-flux calculation developed in the following subsections. Extra details of the calculation are given in Appendix~\ref{App:EnhancedDM density}.

In Ref.~\cite{Hook:2018iia}, authors study the enhanced density of the DM background at the NS conversion
radius. They apply Liouville's theorem to a Maxwell-Boltzmann
velocity distribution defined far from the neutron-star gravitational potential.
Here, we follow the same local phase-space strategy but adapt it to the
angular-momentum-supported core of the minihalo during a NS-PBH inspiral.

We model the local DM velocity distribution in the PBH--core rest frame (far from the neutron star gravitational potential but locally inside the
minihalo core) by an anisotropic Gaussian velocity ellipsoid.
Previous neutron star axion conversion studies 
have adopted either an isotropic Maxwell–Boltzmann distribution for a smooth DM background\,\cite{Hook:2018iia} or an isotropic phase-space distribution reconstructed through Eddington inversion for a spherically symmetric DM spike\,\cite{Edwards:2019tzf}. Both studies then adopt equal radial and tangential velocity dispersions. 

This assumption is less appropriate in the innermost region considered here, where the purely radial infall assumption is no longer valid and the phase-space structure is influenced by angular-momentum support. We therefore allow the radial and tangential velocity dispersions to differ, while retaining a Gaussian distribution normalized to the local core density as the simplest phenomenological generalization of the isotropic Maxwellian. We therefore write
\begin{equation}
f_\infty^{\rm core}(\mathbf w)
=
\frac{\rho_{\rm core}}
{(2\pi)^{3/2}\sigma_r\sigma_t^2}
e^{
-\frac{w_r^2}{2\sigma_r^2}
-\frac{w_\theta^2+w_\phi^2}{2\sigma_t^2}
}\,,
\label{eq:finftycoreW}
\end{equation}
where $\textbf{w}=(w_r, w_\theta, w_\phi)$ is the DM velocity in the PBH--core frame and $\sigma_r$ and $\sigma_t\equiv \sigma_\theta=\sigma_\phi$ are the radial and one-dimensional tangential velocity dispersions, respectively. Since \(r_{\rm core}\) is defined by \(v_{\rm rot}(r_{\rm core})=v_{\rm circ}(r_{\rm core})\), angular-momentum support becomes increasingly important at smaller radii, with \(v_{\rm rot}/v_{\rm circ}\propto(r_{\rm core}/r)^{1/2}\). We therefore adopt a tangentially dominated velocity distribution, \(\sigma_t\gg\sigma_r\), in the inner core.

\begin{figure}[t]
    \centering
    \includegraphics[width=8 cm]{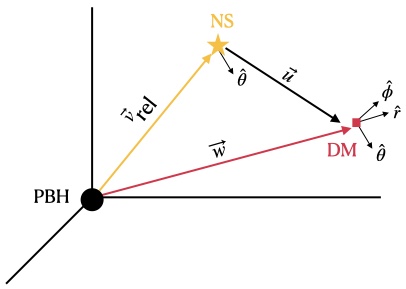}
    \caption{
Schematic definition of the local velocity basis used near the NS orbital position. 
The PBH-centered orthonormal basis \((\hat{\mathbf{r}},\hat{\boldsymbol{\theta}},\hat{\boldsymbol{\phi}})\) is used to decompose both the DM velocity \(\mathbf w\) in the PBH--core frame and the asymptotic DM velocity \(\mathbf u\) in the NS frame. 
For a circular PBH--NS orbit, the NS velocity relative to the core is tangential, $\mathbf v_{\rm rel}=v_{\rm rel}\,\hat{\boldsymbol{\theta}}$, so that
\(\mathbf w=\mathbf u+\mathbf v_{\rm rel}\). 
The arrows \(\mathbf w\) and \(\mathbf u\) represent generic velocity vectors decomposed in this common basis, not additional basis directions. 
 This figure only illustrates the boost between the PBH--core frame and the NS frame; the subsequent gravitational-focusing map from \(\mathbf u\) to the local velocity \(\mathbf{V}_c\) at the conversion radius is treated separately.
}
    \label{fig:framesangles}
\end{figure}

We now move to the NS rest frame. Let
\(\mathbf v_{\rm rel}\) be the velocity of the neutron star with respect to
the PBH--core frame. For a circular PBH--NS orbit, we simply take $\mathbf v_{\rm rel}=v_{\rm rel}\,\hat{\boldsymbol{\theta}}$ where $v^2_\text{rel} = G(M_\text{PBH}+M_\text{NS})/a$, with $a$ as the binary separation. 
We denote
by \(\mathbf u\) the asymptotic dark-matter velocity vector in
the NS frame.
With this convention,
\begin{equation}
\mathbf u=\mathbf w-\mathbf v_{\rm rel}.
\label{eq:boost_convention}
\end{equation}
Therefore the asymptotic velocity distribution written in terms of the incoming
NS--frame variable \(\mathbf u\) is
\begin{equation}
f_\infty^{\rm NS}(\mathbf u)
=
f_\infty^{\rm core}(\mathbf u + \mathbf v_\text{rel}) = f_\infty^{\rm core}(\mathbf w).
\label{eq:fNS_from_fcore}
\end{equation}
Since the transformation is a constant shift in velocity space, the
normalization is unchanged so that both of these distributions integrate to give $\rho_\mrm{core}$.
For $\mathbf v_{\rm rel}=v_{\rm rel}\,\hat{\boldsymbol{\theta}}$, Eq.~\eqref{eq:boost_convention}
gives
\begin{equation}
w_r=u_r,
\qquad
w_\theta=v_{\rm rel}+u_\theta,
\qquad
w_\phi=u_\phi.
\end{equation}
Note that because the Gaussian distribution depends quadratically on \(w_r\) and
\(w_\phi\), the signs of these two components are irrelevant.
A schematic definition of all these velocity variables is shown in
Fig.~\ref{fig:framesangles}. 

Liouville's theorem ensures that the fine-grained phase-space distribution is
conserved along collisionless trajectories. Then, we may equate the velocity distribution function in the NS frame at the conversion radius and at asymptotic infinity $f_c^{\rm NS}(\mathbf x_c,\mathbf V_c)
=
f_\infty^{\rm NS}(\mathbf u)$
where \(\mathbf V_c\) is the actual DM velocity at the conversion
radius. Equivalently, using  Eq.\,\eqref{eq:fNS_from_fcore}, we can write
$
f_c^{\rm NS}(\mathbf x_c,\mathbf V_c)
=
f_\infty^{\rm core}(\mathbf v_{\rm rel}+\mathbf u)$.
The mapping from \(\mathbf V_c\) to the asymptotic velocity is not a simple
Cartesian velocity-space transformation because the angular part of this mapping
is controlled by gravitational focusing and the asymptotic DM velocity distribution is non-isotropic. In the spherical radial-patch
approximation, following the local treatment of Ref.~\cite{Hook:2018iia}, the
shell-averaged density enhancement can be written as
\begin{equation}
\bar\rho_c(r_c)
=
\int \dd^3\mathbf{u}\,
f_\infty^{\rm core}(\mathbf v_{\rm rel}+\mathbf u)
\frac{\sqrt{u^2+v_{\rm esc}^2(r_c)}}{u},
\label{eq:rhocbar_main}
\end{equation}
where $v_{\rm esc}^2(r_c)=2GM_{\rm NS}/r_c$ is
the DM escape velocity at the conversion radius.

Using Eq.~\eqref{eq:finftycoreW} in Eq.~\eqref{eq:rhocbar_main}, we obtain 
\begin{align}
&\bar\rho_c(r_c)
=
\frac{\rho_{\rm core}}
{(2\pi)^{3/2}\sigma_r\sigma_t^2}
\int\dd^3 \textbf{u}
\nonumber\\
&\quad\times
e^{
-\frac{u_r^2}{2\sigma_r^2}
-\frac{(v_{\rm rel}+u_\theta)^2+u_\phi^2}{2\sigma_t^2}
}
\frac{
\sqrt{\textbf{u}^2+v_{\rm esc}^2(r_c)}
}{
|\textbf{u}|
}\,,
\label{eq:rhocint1}
\end{align}
where \(\mathbf u=(u_r,u_\theta,u_\phi)\) is the DM velocity in the NS frame, with components resolved along the local PBH-centered orthonormal basis vectors \((\hat{\mathbf r},\hat{\boldsymbol\theta},\hat{\boldsymbol\phi})\).
This is the boosted anisotropic analogue of the density-enhancement integral
used for an isotropic dark-matter background.

We now take the regime relevant for the inner angular-momentum-supported core. Using Eqs.\,\eqref{eq:vrot} and \,\eqref{eq:rcore} for the DM tangential velocity at asymptotic infinity, $v^2_\text{rot} \sim 2\sigma^2_t$, we have that  
\begin{equation}
\frac{v_\text{rot}}{v_\text{rel}} = \left(\frac{M_\text{PBH}}{M_\text{PBH}+M_\text{NS}} \frac{r_\text{core}}{r} \right)^{1/2}\,.
\end{equation}
For $M_{\rm PBH}=(1\text{--}10)\,M_\odot$, the characteristic tangential DM speed is comparable to the NS–PBH relative speed near \(r_{\rm core}\) and dominates at smaller radii. Therefore, we cannot assume  $\sigma_t \gg v_\text{rel}$ in the whole core. Thus, keeping the full dependence on $ v_\text{rel}/\sigma_t$ and only using $\sigma_r \ll \sigma_t$, Eq.~\eqref{eq:rhocint1}
reduces to  
\begin{align}
\bar\rho_c(r_c)
\simeq
\frac{\rho_{\rm core}}{\sigma_t^2}&e^{-\frac{v^2_\text{rel}}{2\sigma_t^2}}
\int_0^\infty \dd u_\perp\,
e^{-\frac{u_\perp^2}{2\sigma_t^2}}\times \nonumber\\
&I_0\left( \frac{u_\perp v_\text{rel}}{\sigma_t^2} \right)
\sqrt{u_\perp^2+v_{\rm esc}^2(r_c)}\,,
\label{eq:mainrhoc_tangential_integral}
\end{align}
where $I_0(z)$ is the modified Bessel function of the first kind of zero order and $u_\perp =
(u^2_\theta + u^2_\phi)^{1/2}$ is the DM speed perpendicular to the radial component.
Depending on the neutron-star parameters and the axion mass, the conversion
radius may lie sufficiently close to the stellar surface that
\(v_{\rm esc}\gg \sigma_t, v_{\rm rel}\). In that limit, Eq.~\eqref{eq:mainrhoc_tangential_integral}
reduces to
\begin{equation}
\bar\rho_c(r_c)
\simeq
\rho_{\rm core}\sqrt{\pi}\,
\frac{v_{\rm esc}(r_c)}{v_{\rm rot}}e^{-\frac{v^2_\text{rel}}{2v^2_\text{rot}}}I_0\left( 
\frac{v^2_\text{rel}}{2v^2_\text{rot}}\right).
\label{eq:rhoc_strong_focusing}
\end{equation}
where we have used $v^2_\text{rot} \simeq 2 \sigma_t^2$. 

As a consistency check, in the additional limit 
$v_\text{rel} \ll v_\text{rot}$, meaning the tangential DM velocity dominates over the NS–PBH relative velocity, Eq.\,\eqref{eq:rhoc_strong_focusing} simplifies to
\begin{equation}
\bar\rho_c(r_c)
\simeq
\rho_{\rm core}\sqrt{\pi}\,
\frac{v_{\rm esc}(r_c)}{v_{\rm rot}}.
\end{equation}
This expression has the same parametric structure as the standard DM-background
result of Ref.~\cite{Hook:2018iia}, up to a numerical factor of order unity,
under the replacement
\(\rho_{\rm DM}^{\infty}\to\rho_{\rm core}\) and
\(v_0\to v_{\rm rot}\). In the following, we denote the spherical radial-patch
estimate \(\bar\rho_c\) simply by \(\rho_c\).

\subsubsection{Axion-photon conversion probability and Radiation power}
\label{subsec:RadiatedPower}

Having specified the axion phase-space distribution and obtained the corresponding local axion velocity and density at the conversion radius, we now calculate the axion--photon conversion probability in the neutron-star magnetosphere and the associated radiated power. We begin from the axion--photon interaction term in the Lagrangian-density,
\begin{equation}
\mathcal{L}_{\rm int}
=
-\frac{1}{4}g_{a\gamma\gamma}\,a\,F_{\mu\nu}\tilde F^{\mu\nu}
=
g_{a\gamma\gamma}\,a\,\mathbf{E}\cdot\mathbf{B},
\end{equation}
where $a(\mathbf{x},t)$ is the axion field and  \(g_{a\gamma\gamma}\) is the axion--photon coupling constant. In the presence of the NS magnetospheric magnetic field, this interaction
allows mixing between the axion field and the photon polarization that is
transverse to the axion trajectory and coplanar with the magnetic field.
Equivalently, the mixing is controlled by the component
\(B_t = |\mathbf{B}|\sin\tilde\theta\), where \(\tilde\theta\) is the angle
between the axion propagation direction and the local magnetic field. Assuming a radial axion trajectory at the conversion radius and using the WKB approximation, together with the stationary-phase method, the axion--photon conversion probability at infinity takes the analytic form\,\cite{Hook:2018iia}
\begin{equation}
p_{a\gamma}^{\infty}(\tilde\theta)
\simeq
\frac{\pi}{3}\,
g_{a\gamma\gamma}^{\,2}
\frac{
 B^2\!\left( r_c\right)\,  r_c
}{
m_a
}
\left[
\frac{
1+V_c^2
}{
1+V_c^2\sin^2\tilde\theta
}
\right]^{2/3}\,,
\label{eq:probgeneral_tildetheta}
\end{equation}
where, using \cref{eq:rcfull,eq:Bgeneral}, the squared magnetic-field magnitude is
\begin{equation}
\begin{split}
B^2(&r_c,\theta,\theta_m,t)=\frac{B^2_0}{4}\left(  \frac{R_\text{NS}}{r_c}\right)^6 \\
&\times\left[ 1 + 3(\text{cos}\theta_m\text{cos}\,\theta + \text{sin}\theta_m\, \text{sin}\theta\, \text{cos}(\Omega t))^2\right]\,,\\
\end{split}
\end{equation}
and the conversion radius is $r_c(\theta,\theta_m,t)=r_\star |A_1 + A_2\text{cos}(\Omega t)|^{1/3}$
with 
\begin{align}
r_\ast
&\equiv
R_{\rm NS}
\left(
\frac{\omega_{{\rm pl},0}}{m_a}
\right)^{2/3},\label{eq:rstar}
\\
A_1
&\equiv
\cos\theta_m\,
\left(3\cos^2\theta-1\right),
\\
A_2
&\equiv
3\sin\theta_m\sin\theta\cos\theta .
\label{eq:main_A_C}
\end{align}

Expanding \cref{eq:probgeneral_tildetheta} in the non-relativistic regime \(V_c\ll 1\) , we find
\begin{equation}
\begin{split}
p_{a\gamma}^{\infty}(\tilde\theta)
&\simeq
\frac{\pi}{3}\,
g_{a\gamma\gamma}^{\,2}
\frac{
 B^2\!\left(r_c\right)\, r_c
}{
m_a
}\\
&\times
\left[
1+\frac{2}{3}V_c^2\cos^2\tilde\theta
+\mathcal{O}(V_c^4)
\right]\,,
\end{split}
\label{eq:pag_general_tildetheta_expanded_explicit}
\end{equation}
where the leading order correction is quadratic in $V_c$. The actual angle between the magnetic field and the radial direction can be readily calculated as $\text{cos}\,\tilde\theta = \hat{\textbf{B}}\cdot \hat{\textbf{r}}$. Explicitly, we have 
\begin{equation}
\!\!\text{cos}\,\tilde \theta = \frac{2[\text{cos}\theta_m\, \text{cos}\theta+\text{sin}\theta_m \,\text{sin}\theta\, \text{cos}(\Omega t)]}{\sqrt{1+3(\text{cos}\theta_m\, \text{cos}\theta + \text{sin}\theta_m\, \text{sin}\theta \,\text{cos}(\Omega t))^2}}\,.
\end{equation}
The local axion mass flux incident on the conversion surface within the neutron-star magnetosphere is $\Phi_{\rm in}^{(c)}
\simeq
\bar\rho_{\rm c}(r_c) V_c $. Then, the differential radiated power per unit of solid angle on the conversion surface is estimated as
\begin{equation}
\frac{\dd\mathcal{P}}{\dd\Omega} \simeq  2r_c^2 \Phi_{\rm in}^{(c)} p_{a\gamma}^{\infty} \,,
\end{equation}
where the prefactor of two accounts for the possibility of axion–photon conversion during either the inward or outward crossing of the conversion shell, following Ref.\,\cite{Hook:2018iia}. The incoming and outgoing branches entering the calculation of the local DM density are already included in \(\bar\rho_c(r_c)\), as discussed in Appendix~\ref{App:EnhancedDM density}.

\subsubsection{Time-averaged Spectral Flux Density}
\label{subsec:SpectralFluxDensity}

Consider an NS source located at a distance $d$ from the Earth. For a normalized spectral line profile $p(\nu)$, the measured spectral flux density takes the form 
\begin{equation}
S_\nu(\theta,\theta_m,t;\nu) = \funop{F(\theta,\theta_m,t)}\funop{p(\nu)}\,,
\label{eq:tophatprofile}
\end{equation}
where $F(\theta,\theta_m,t) = (\dd\mathcal{P}/d\Omega)/d^2$ is the radio flux at Earth associated with the axion-induced spectral line.
For a normalized top-hat profile of width \(\Delta\nu\),
such that \(p(\nu)=1/\Delta\nu\), we recover \(S_\nu=F/\Delta\nu\), as used in
Refs.\,\cite{Hook:2018iia, Edwards:2019tzf}. The characteristic width of the line receives contributions from the intrinsic axion velocity dispersion, $\Delta\nu_{\rm int}$, the bulk Doppler motion of the neutron star, $\Delta\nu_{\rm bulk}$,
and the magnetospheric co-rotation broadening, $\Delta\nu_{\rm corot}$.
Therefore, the total bandwidth may be estimated by combining these contributions in quadrature, 
\begin{equation}
\Delta\nu^2
\simeq
\Delta\nu_{\rm int}^2
+
\Delta\nu_{\rm bulk}^2
+
\Delta\nu_{\rm corot}^2\,.
\label{eq:main_Dnu_eff}
\vspace{0.4cm}
\end{equation}
The relative size  of the different  widths depends on
the neutron-star properties, the axion mass, and the source geometry. 

First, the intrinsic axion-dispersion width is set by the spread in the
nonrelativistic kinetic energy of the incoming axion population in the NS
frame. With the convention introduced in Sec.\,\ref{subsec:DM density}, we have
\begin{equation}
\nu_a
\simeq
\frac{m_a}{2\pi}
\left(1+\frac{u^2}{2}\right)\,.
\label{eq:nu_axion_kinetic}
\end{equation}
The intrinsic distribution of $\mathbf{u}$ comes from the PBH--core frame distribution through the relative velocity $\mathbf{w}$, introduced in \cref{eq:boost_convention}. The characteristic velocity spread entering the intrinsic linewidth is
the spread of the NS-frame incoming velocity distribution. Thus,
\begin{equation}
\Delta\nu_{\rm int}
\sim
\nu_{\rm peak}\,\langle u_\perp^2\rangle
\simeq
\frac{m_a}{2\pi}\,v_\text{rot}^2 ,
\label{eq:Dnuint_tangential}
\end{equation}
where we have used  $u^2\simeq
u_\perp^2 \simeq  v_\text{rot}^2$ at $r\ll r_\text{core}$ and
\(\nu_{\rm peak}\simeq m_a/(2\pi)\).

Second, the bulk kinematic width arises from the orbital motion of the emitting
neutron-star magnetosphere relative to the observer. It corresponds to the
ordinary Doppler broadening of the line center,
\begin{equation}
\Delta\nu_{\rm bulk}
\sim
\nu_{\rm peak}\,
v_{\rm NS}\,,
\label{eq:nubulk}
\end{equation}
up to projection factors and the orbital phase interval covered by the
observation. Here $v_{\rm NS}$ is the NS orbital speed around the binary center of mass as was already defined in Sec.\,\ref{sec:gw_sig}.

Third, for misaligned neutron stars, we have the additional contribution of the co-rotation broadening. Such a broadening scales as~\cite{Battye:2021xvt, Battye:2019aco}
\[
\Delta\nu_{\rm corot}\sim \nu_{\rm peak}\,r_\star\,\Omega\sin\theta_m,
\]
where \(r_\star\) is the characteristic resonant conversion radius, Eq.\,(\ref{eq:rstar}), \(\Omega=2\pi/P\) is the stellar spin frequency, and \(\theta_m\) is the magnetic obliquity as explained before. For  $r \ll r_\text{core}$, for the NS properties displayed in Table\,\ref{tab:NStargets} and axion masses in the range of interest for radio telescopes, the dominant bandwidth source comes from the NS orbital motion and the intrinsic axion-dispersion, so that $\Delta\nu \sim \sqrt{\Delta\nu^2_{\rm bulk}+\Delta\nu_{\rm int}^2}$. This bandwidth feature departs from the standard case of DM background falling into an isolated NS magnetosphere, where the broadening signal source comes from $\Delta\nu_{\rm int}$ or $\Delta\nu_{\rm corot}$, for aligned or misaligned NSs, respectively\,\cite{Leroy:2019ghm}. The bulk Doppler bandwidth for a quasi-circular binary with an initial orbital separation $a_0$ can be written as
\begin{equation}
\Delta\nu_\text{bulk} \sim \frac{m_a}{2\pi}\frac{M_\text{PBH}}{M_\text{tot}}\sqrt{\frac{GM_\text{tot}}{r(t)}}\,,
\label{eq:deltanudominant}
\end{equation}
 where \(r(t)\) is determined from the orbital energy-balance equation, Eq.\,\eqref{eq:r_dot}. 

For the radio observations relevant for our estimates, the integration time per
pointing or per image is typically longer than the spin period of the neutron
stars considered here. For instance, SKA pulsar-survey forecasts often adopt
integration times of order \(30\,{\rm min}\)\,\cite{Smits:2008cf}, 
searches of nearby isolated NSs via the GBT performed by authors in Ref.\,\cite{Foster:2020pgt} used \(20\,{\rm min}\), and the archival
VLA transient search of Bell et al.~accumulated a total observing time of
\(435\,{\rm h}\), with an average integration time per image of
\(5.2\,{\rm min}\)
\cite{2011MNRAS.415....2B}.
For benchmark neutron-star periods \(P\sim{\cal O}(1-10)\,{\rm s}\), these
observing strategies satisfy $t_{\rm obs}\gg P$. 

The phase-resolved signal is therefore averaged over many stellar rotations,
and the relevant observable is the spin-averaged spectral flux density,
\begin{equation}
\left\langle S_\nu\right\rangle_{\rm spin}
=
\left\langle F\right\rangle_{\rm spin}\,p(\nu)\,,
\label{eq:Sspin}
\end{equation}
where
\begin{align}
\left\langle F\right\rangle_{\rm spin}
&=
\frac{1}{P}
\int_0^P \dd t\,
\frac{1}{d^2}
\frac{\dd\mathcal{P}(t)}{\dd\Omega} 
\,\nonumber\\
&\times\Theta\!\left[
\left|A_1+A_2\cos(\Omega t)\right|
-
\left(
\frac{m_a}{\omega_{{\rm pl},0}}
\right)^2
\right].
\end{align}
The Heaviside step function above ensures that 
the axion-photon conversion occurs in the NS magnetosphere, $r_c > R_\text{NS}$. The time dependence of the spectral flux density is encoded in the magnetic field, the conversion radius, and the signal bandwidth.

\subsubsection{Sensitivity to nearby neutron stars}
\label{subsec:Sensitivity}

As a benchmark facility, we consider the Green Bank Telescope (GBT) as a representative
single-dish radio telescope for our sensitivity projections.
\begin{table}[t]
\centering
\renewcommand{\arraystretch}{1.15}
\begin{tabular}{lccc}
\hline\hline
Receiver / band 
& SEFD frequency interval 
& $B_{\rm inst}^{\rm eff}$ 
& $n_{\rm pol}$ \\
& [GHz] 
&  [GHz]
&  \\
\hline
PF1 342 MHz
& $0.290$--$0.395$
& $0.105$
& $2$ \\

PF1 450 MHz
& $0.385$--$0.520$
& $0.135$
& $2$ \\

PF1 600 MHz
& $0.510$--$0.690$
& $0.180$
& $2$ \\

PF1 800 MHz
& $0.680$--$0.920$
& $0.240$
& $2$ \\

PF2
& $0.910$--$1.230$
& $0.320$
& $2$ \\

L band
& $1.15$--$1.73$
& $0.580$
& $2$ \\

S band
& $1.73$--$2.60$
& $0.870$
& $2$ \\

\hline\hline
\end{tabular}
\caption{
GBT receiver bands used in the baseline spectral-line sensitivity estimate. The frequency intervals in the second column are restricted to the receiver ranges over which the SEFD is shown in Fig.~4 of the 2026
GBT Proposer's Guide\,\cite{GBTProposersGuide2026}. When the GBT guide lists the instantaneous bandwidth as ``full'', we set \(B_{\rm inst}^{\rm eff}\) equal to the width of the SEFD-supported frequency interval used in our sensitivity curve. Therefore, small differences between broader tunable ranges and the SEFD intervals are not extrapolated. For example, for S band we use
\(1.73\)--\(2.60\,{\rm GHz}\) and
\(B_{\rm inst}^{\rm eff}=0.870\,{\rm GHz}\), rather than the broader tunable range \(1.680\)--\(2.650\,{\rm GHz}\). 
}
\label{tab:gbt_receivers_sefd}
\end{table}
We estimate the radio sensitivity using the radiometer equation. For a
flux density $\langle S_\nu \rangle_\text{spin}$, Eq.\,\eqref{eq:Sspin}, the root-mean-square (rms) noise is
\begin{equation}
\sigma_S =
\frac{{\rm SEFD}}
{\sqrt{n_{\rm pol} B t_{\rm obs}}}\,,
\label{eq:sigmas}
\end{equation}
where \({\rm SEFD}\) is the system equivalent flux density,
\(n_{\rm pol}\) is the number of summed polarizations, \(B\) is the
frequency bandwidth, and \(t_{\rm obs}\) is the integration time. The GBT receiver bands, effective instantaneous bandwidth ($B^\text{eff}_\text{inst}$) used in the sensitivity estimate are listed in Table \,\ref{tab:gbt_receivers_sefd}. The effective instantaneous bandwidth per tunnable frequency receivers is set to match the width of the SEFD curve with typical galactic background shown in Fig.~4 of Ref.\,\cite{GBTProposersGuide2026}. 

For our sensitivity estimates, we adopt a quasi-static snapshot at \(r_0=r(f_{\rm GW}=0.1\,{\rm Hz})\). The focused axion density and all orbital-dependent linewidths are evaluated at \(r_0\) and held fixed during \(t_{\rm obs}=1\) and \(10\) hr.
The conversion radius entering in the velocity escape calculation is estimated by evaluating it at \(t=0\,\text{s} \).
Appendix~\ref{sec:radio_integration_time} shows that their orbital evolution is negligible over these integration times. 

Therefore, the signal-to-noise ratio takes the form
\begin{equation}
{\rm SNR}^2 =
\frac{n_{\rm pol}t_{\rm obs}}{{\rm SEFD}^2(\nu_{\rm peak})}
\int_{\nu_-}^{\nu_+}\dd \nu \, \left\langle S_\nu\right\rangle_{\rm spin}^2,
\label{eq:SNR}
\end{equation}
where \([\nu_-,\nu_+]\) are the lower and upper
edges of the SEFD-supported frequency interval of receiver \(R\), respectively, listed
in Table\,\ref{tab:gbt_receivers_sefd}\footnote{Here the bandwidth factor \(B\) appearing in the standard radiometer equation is incorporated explicitly through the frequency integral. As a simple check, if $S_\nu = S$ is constant over a bandwidth \(B\), then Eq.\,\eqref{eq:SNR} gives $\text{SNR} = (\langle S \rangle/\text{SEFD})\sqrt{n_\text{pol}Bt_\text{obs}}$, in agreement with Eq.\,\eqref{eq:sigmas}.}. We take $p(\nu)$ to be a Gaussian with mean $\nu_\mrm{peak} \simeq  m_a/(2\pi)$ and variance $\Delta \nu^2$. 

Since $\langle F(\nu_\text{peak})\rangle_{\rm spin} \propto g^2_{a\gamma\gamma}$, 
we write
\begin{equation}
\langle F\rangle_{\rm spin}=\langle F\rangle^{\rm ref}_{\rm spin}\left( \frac{g_{\phi\gamma\gamma}}{g^{\text{ref}}_{\phi\gamma\gamma}}\right)^2\,,
\end{equation}
so that the minimum detectable coupling can be written as
\begin{equation}
\begin{split}
g^\text{min}_{a\gamma\gamma}= 
g^\text{ref}_{a\gamma\gamma}&\left[ \frac{\text{SNR}_\text{th}\text{SEFD}(\nu_\text{peak})}{\langle F\rangle^{\rm ref}_{\rm spin}}\right]^{1/2}\\
\times
&\left[ \frac{4\sqrt{\pi}\Delta\nu}{n_\text{pol}t_\text{obs}\varepsilon(\nu_\text{peak})}\right]^{1/4}\,,
\label{eq:gmin}
\end{split}
\end{equation}
where $\text{SNR}_\text{th} = 5$ is the chosen detection threshold for the signal-to-noise ratio  and
\begin{equation}
\varepsilon(\nu_\text{peak}) = \text{erf}\left( \frac{\nu_+-\nu_\text{peak}}{\Delta\nu} \right)-
\text{erf}\left( \frac{\nu_--\nu_\text{peak}}{\Delta\nu} \right)\,.
\end{equation}
\begin{figure}
    \centering
    \includegraphics[width=0.477\textwidth]{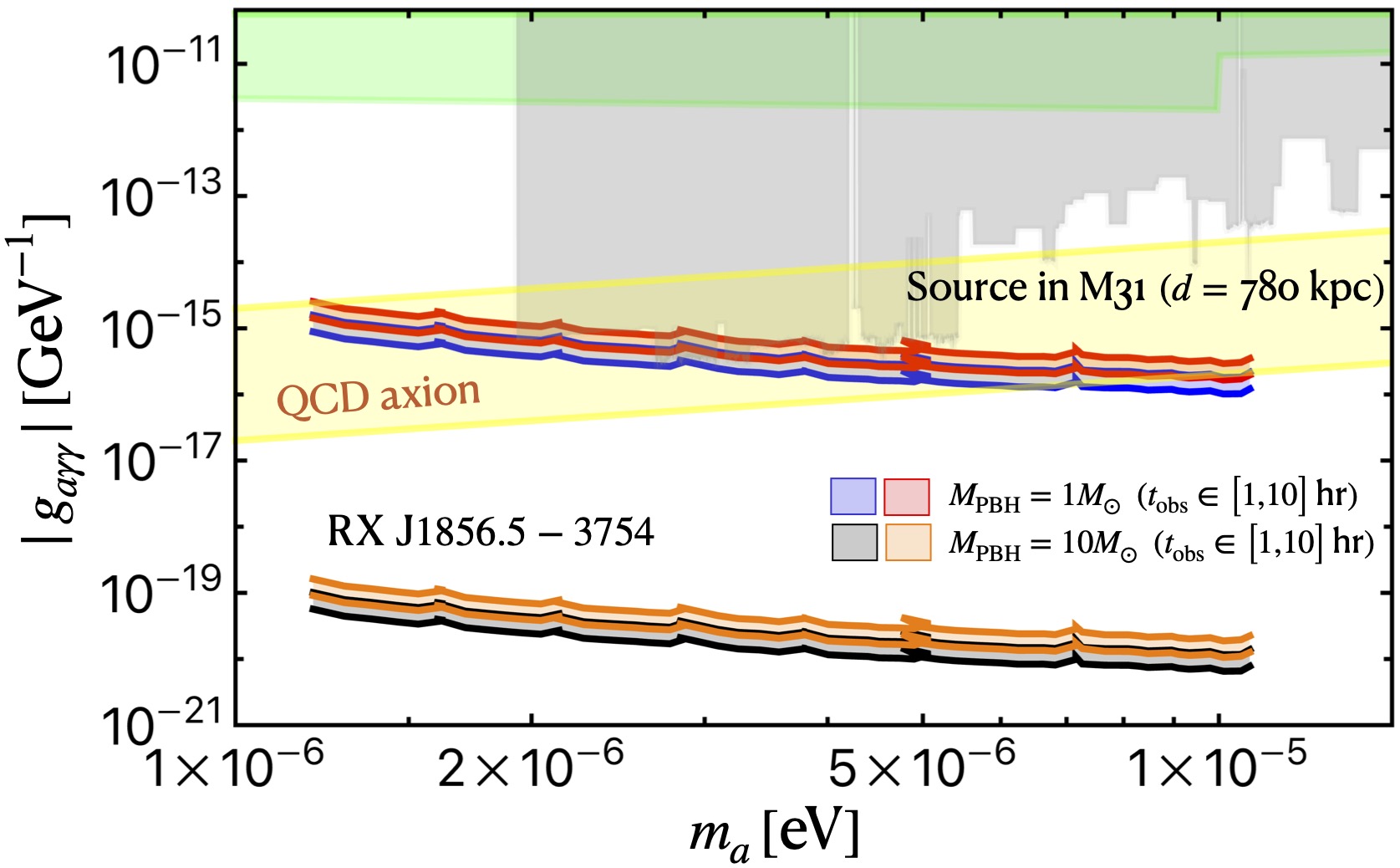}
    \caption{Sensitivity plot for the axion-photon coupling $g_{a\gamma\gamma}$ assuming a signal-to-noise ratio of $5$. Two sources are considered: the nearby isolated NS RX J1856.5-3754 ($\theta = 10^{\text{o}}$) and an extragalactic-distance source benchmark ($\theta = 10^{\text{o}}$) placed at M31 (astrophysical properties listed in Table~\ref{tab:NStargets}). The colored band indicates an observation time between 1 and 10 hours using the GBT one-dish telescope. While the green shaded region indicate constraints from pulsar polar-cap Cascades\,\cite{Noordhuis:2022ljw} and an axion helioscope\,\cite{Ruz:2024gkl}, gray shaded region refers to haloscope constraints\,\cite{ADMX:2025vom,ADMX:2021nhd,ADMX:2018gho,PhysRevLett.124.101303,ADMX:2021mio,ADMX:2018ogs,Adair:2022rtw,CAPP:2024dtx,PhysRevLett.59.839,Alesini:2019ajt,Yi:2022fmn,Jeong:2020cwz,HAYSTAC:2024jch}. The yellow band shows the QCD axion parameter of space, adapted from AxionLimits \cite{AxionLimits}.}
    \label{fig:sensitivity}
\end{figure}
\begin{table}[t]
\centering
\caption{Envelope of the projected coupling reach for the neutron-star benchmarks in Table~\ref{tab:NStargets}, assuming \(t_{\rm obs}=10\,{\rm h}\) and \(M_{\rm PBH}=1\,M_\odot\). The angle \(\theta\) is the polar angle used in the plasma-frequency calculation.}
\label{tab:gmin_envelope}
\scriptsize
\setlength{\tabcolsep}{3.5pt}
\renewcommand{\arraystretch}{1.15}
\resizebox{\columnwidth}{!}{%
\begin{tabular}{lcc}
\hline\hline
Source & \(\theta\) 
& PF1--S \\
& 
& \(g_{\rm min}\,[{\rm GeV}^{-1}]\) \\
\hline
RX~J1856.5$-$3754
& \(10^\circ\)
& \(7\times10^{-21}\)--\(6\times10^{-20}\) \\

RX~J0806.4$-$4123
& \(10^\circ\)
& \(2\times10^{-20}\)--\(1\times10^{-19}\) \\

RX~J0720.4$-$3125
& \(0^\circ\)
& \(2\times10^{-20}\)--\(1\times10^{-19}\) \\

Typical galactic NS
& \(20^\circ\)
& \(1\times10^{-19}\)--\(1\times10^{-18}\)\\

3XMM~J004232.1$+$411314
& \(10^\circ\)
& \(2\times10^{-16}\)--\(1\times10^{-15}\)\\
\hline\hline
\end{tabular}%
}
\end{table}
\cref{fig:sensitivity} shows the projected sensitivity to the axion--photon coupling
$g_{a\gamma\gamma}$ obtained with GBT for two representative neutron-star targets. 
The first is the nearby isolated neutron star RX~J1856.5--3754, which we use as a clean 
representative member of the Magnificent Seven. The second is an idealized extragalactic 
benchmark motivated by 3XMM~J004232.1+411314, a known accreting X-ray pulsar in M31.
For the latter case, we use the observed distance and spin period only as motivation for an 
M31-distance neutron-star benchmark, while the magnetic field and Goldreich--Julian plasma 
profile should be understood as idealized inputs.

The projected curves cover the GBT frequency range from the PF1 band up to the S band, corresponding to the axion masses $m_a \simeq  (10^{-6} - 10^{-5})\,{\rm eV}$
or, equivalently, to the radio frequencies $\nu_a \simeq (0.29\text{--}2.60)\,{\rm GHz}$. Within this interval, the resonant conversion radius remains outside the neutron
star for the benchmark cases entering Table~\ref{tab:NStargets}. The
typical galactic neutron-star benchmark can in principle probe somewhat higher
frequencies, up to $\nu_a\sim 10~{\rm GHz}$, corresponding to
$m_a\simeq 4.1\times 10^{-5}~{\rm eV}$, because its shorter spin period and
fiducial magnetic field allow the magnetospheric plasma frequency to match
larger axion masses. However, this higher-frequency extension is not included in
Table~\ref{tab:gmin_envelope}, which is intended as a conservative comparison
over the common PF1--S range.

For each target we consider observation times of $t_{\rm obs}=1$ and $10\,{\rm hr}$, adopting the quasi-static snapshot at \(r_0=r(f_{\rm GW}=0.1\,{\rm Hz})\) described above. As expected, increasing either the central PBH mass or the 
integration time improves the sensitivity. This follows directly from Eq.~\eqref{eq:gmin}. 
In the Doppler-broadened regime, and for the inner UCMH core density used in this work, 
one has schematically
\[
g_{\phi\gamma\gamma}^{\rm min}
\propto 
\rho_{\rm core}^{-1/2}\,
\Delta\nu^{1/4}\,
t_{\rm obs}^{-1/4}
\propto 
M_{\rm PBH}^{-1/8}\,
t_{\rm obs}^{-1/4}.
\]
Thus, the gain with observing time is real but slow. Similarly, the dependence on the PBH mass is mild because the 
increase in the UCMH core density is partly compensated by the corresponding increase in 
the Doppler width.

The full envelope of the projected coupling reach for all neutron-star benchmarks listed in 
Table~\ref{tab:NStargets} is summarized in Table~\ref{tab:gmin_envelope}. For galactic sources, especially nearby isolated
Magnificent Seven targets, GBT is projected to be sensitive to couplings at least two orders of magnitude weaker than the QCD-axion band in the mass range
covered by the receiver bands we have considered. This indicates that NS--PBH
inspirals inside QCD-axion UCMHs could test genuine QCD-axion parameter space,
rather than only generic ALP scenarios. For the M31-distance benchmark, the
larger distance produces a severe flux suppression, so the sensitivity is much
weaker. However, the projected spectral flux density is still strong enough to place the minimal required axion-photon coupling constant for detection within the QCD axion band, going below it for the most favorable assumptions---larger
$M_{\rm PBH}$, longer integration time, and the higher-frequency side of the
considered axion-mass interval.

\section{Conclusions}

We have studied the multimessenger signatures arising from a neutron star inspiralling into a primordial black hole in the presence of a QCD-axion ultracompact minihalo. We have considered a mixed dark-matter scenario in which axions dominate the particle component and PBHs of $M_\mrm{PBH} = (1\text{--}10)\,M_\odot$ form a subdominant compact component that seeds the formation of these structures. Our main results are the following.

\begin{itemize}[leftmargin=1.5em]
    \item Starting from pre-inflationary Peccei–Quinn initial conditions consistent with relic-abundance and isocurvature bounds, we used the secondary-infall framework to reproduce the characteristic $\rho_\mrm{halo} \propto M_\mrm{PBH}^{3/4} r^{-9/4}$ profile, finding a fitted slope of $2.24$. The evolution is controlled almost entirely by the central PBH seed, with axion isocurvature perturbations and the initial peculiar motion of the accreting shells yielding negligible corrections. In the innermost region of the ultracompact minihalo, we adopted a constant density core as a deliberately minimally concentrated benchmark, making our signal estimates conservative.

    \item Dynamical friction due to the presence of the minihalo adds an orbital-energy loss channel on top  of the usual emission of gravitational waves, dephasing the gravitational wave waveform relative to the vacuum scenario. Comparing the reconstructed signal-to-noise ratios from a matched filter analysis when using the correct dressed waveform against the vacuum template, this dephasing remains distinguishable across the LISA band over our considered PBH mass range. Gravitational waves therefore carry the signature of the presence of ultracompact minihalos seeded by stellar mass PBHs, and the future gravitational wave interferometer LISA will be able to detect them.

    \item Resonant axion–photon conversion in the neutron-star magnetosphere probes the particle content of the same halo. We include gravitational focusing of the axion phase-space distribution and the dominant broadening mechanism from the bulk kinematic width. For a Goldreich–Julian magnetosphere and Green Bank Telescope observations over the PF1–S range with $t_\mrm{obs}=1\text{--}10\,\mathrm{hr}$, the projected $g_{a\gamma\gamma}$ sensitivity for nearby galactic targets, for which we consider the Magnificent Seven in particular, lies at least two orders of magnitude below the QCD band for $\mathrm{\mu eV}$ masses. Even an Andromeda source reaches into the band despite the flux suppression. The reach improves with longer integration, heavier central PBHs, and higher frequency.
\end{itemize}

The central result of this work is the complementarity of the two messenger signals. Gravitational waves are sensitive to the density of the dark-matter structure surrounding the PBH, whereas the radio counterpart directly probes the axion component through its coupling to photons. A coincident observation will consequently test, within a single astrophysical system, the presence of a compact primordial object, the dense dark-matter structure assembled around it, and the particle nature of that dark matter. This provides a qualitatively different search strategy from experiments separately targeting either PBHs or axions, illustrating how mixed dark sectors containing both compact and particle dark matter can lead to distinctive multimessenger phenomena. Improving the modeling of the angular-momentum-supported inner halo, the axion phase-space distribution, realistic neutron-star magnetospheres, and the astrophysical survival and encounter rates of these systems will be important steps toward translating this possibility into a dedicated multimessenger search strategy.

\section{Acknowledgments}
EDS truly thanks Tsutomu Yanagida (University of Tokyo), Kimmo Kanulainen (University of Jyväskylä), and Sami Nurmi (University of Jyväskylä and University of Helsinki) for enriching discussions in the early stages of this work. This research was supported by FONDECYT Project N° 1251141 (ANID, Chile). 
EDS is grateful to Director Jose Villanueva for his hospitality during a research stay at the Institute of Physics and Astronomy (IFA), Universidad de Valparaíso, in December 2025. Additionally, EDS thanks the Institut de Física d'Altes Energies (IFAE) at the Universitat de Barcelona for their hospitality during a visit in February 2026, which facilitated key discussions with Diego Blas and the co-author Dorian Amaral. Both research visits were made possible through the support of the aforementioned Fondecyt grant. DA has been supported by ERC grant ERC-2024-SYG 101167211 by the European Union.
Views and opinions expressed are however those of the author(s) only and do not necessarily reflect those of the European Union, European Research Council Executive Agency, or other awarding body. Neither the European Union nor the granting authority can be held responsible for them.
\\

\hspace{-0.35 cm}
$*$\href{mailto:Enrico.Schiappacasse@uss.cl}{Enrico.Schiappacasse@uss.cl}
(corresponding author)

\appendix

\section{Fraction of isolated PBHs that acquire minihalos}
\label{app:fractionisolated}

For simplicity, we consider a monochromatic mass function for PBHs of mass $M_{\text{PBH}} \sim 10 M_{\odot}$. We adopt the notation used in Ref.~\cite{Sasaki:2016jop}. The physical mean separation between two PBHs at the redshift of matter-radiation equality, $z_{\text{eq}}$, is defined as
\begin{equation}
\begin{split}
\overline{x} &= \left( \frac{3 M_{\text{PBH}}}{4\pi \rho_{\text{PBH}}(z_{\text{eq}})} \right)^{1/3}\\
&= \left( \frac{3M_{\text{PBH}}}{4\pi (1+z_{\text{eq}})^3 f_{\text{PBH}}\Omega_{\text{DM}}\rho_{c,0}} \right)^{1/3} \\
&\simeq 2\, \text{pc} \left( \frac{10^{-3}}{f_{\text{PBH}}} \right)^{1/3} \left( \frac{M_{\text{PBH}}}{30 M_{\odot}} \right)^{1/3}\,,
\end{split}
\label{rbara}
\end{equation}
where $\rho_{\text{PBH}}(z)$ is the mean PBH energy density at a given redshift. Consider two neighboring PBHs separated by a physical distance  $x$ at $z_{\text{eq}}$, such that $x \leq \overline{x}$. Take $z_{\text{dec}}$ as the redshift at which the pair of PBHs break free from the Hubble flow and form a gravitationally bound system. The binary formation occurs when the local energy density of the pair, $\rho_{\text{pair}}(z_{\text{dec}})$, exceeds the background radiation energy density, $\rho_r(z_{\text{dec}})$. Specifically,
\begin{align}
\rho_{\text{pair}}(z_{\text{dec}})& = \frac{3 M_{\text{PBH}}}{4\pi x^3} > \rho_r(z_{\text{dec}})\,,\\
\rho_{\text{PBH}}(z_{\text{dec}})\left( \frac{\overline{x}}{x}\right)^{3} & > \frac{\rho_{\text{eq}}}{2}\left( \frac{1+z_{\text{dec}}}{1+z_{\text{eq}}} \right)^4\,,\\
f_{\text{PBH}}\left( \frac{\overline{x}}{x} \right)^3 &> \left( \frac{1+z_{\text{dec}}}{1+z_{\text{eq}}}\right) \,,\label{fxx}
\end{align}
where we have used 
$\rho_{\text{PBH}}(z_\mrm{dec})=\rho_{\text{DM}}(z_\mrm{dec}) f_{\text{PBH}} \approx (\rho_{\text{eq}}/2)[(1+z_{\text{dec}})/(1+z_{\text{eq}})]^3f_{\text{PBH}}$ in the third line. Note that if $f_{\text{PBH}}^{1/3} (\overline{x}/x) < 1$,
 we have $\rho_{\text{pair}}(z) = \rho_{\text{PBH}}(z)(\overline{x}/x)^3 = \rho_{\text{DM}}(z)f_{\text{PBH}}(\overline{x}/x)^3 < \rho_{\text{PBH}}(z)$. The pair of PBHs will never decouple from the Universe expansion. Thus, the decoupling should take place before the matter-radiation equality, i.e. $(1+z_{\text{dec}}) > (1+ z_{\text{eq}})$~\cite{Sasaki:2016jop}.
 
 At the time of the binary formation, the semi-major axis can be expressed as a function of the distances $x$ and $\overline{x}$, and a proportionality constant $\alpha$ by using Eq.~(\ref{fxx}) as
 \begin{equation}
a =\alpha x\frac{(1+z_{\text{eq}})}{(1+z_{\text{dec}})} =  \frac{\alpha x^4}{\overline{x}^3 f_{\text{PBH}}}\,.     
 \end{equation}
 Within the Newtonian approximation and treating PBHs as point particles, numerical studies suggest that $\alpha = 0.4$~\cite{Ioka:1998nz, Ali-Haimoud:2017rtz}. Taking the decoupling time as $z_{\text{dec}}\simeq z_{\text{eq}}$, the maximum semi-major axis for binaries is estimated to be $a_{\text{max}} = \alpha f_{\text{PBH}}^{1/3}\overline{x}$. 

In the three body approximation, tidal forces acting on the PBH pair coming from the closest PBH provides the needed angular momentum to avoid a head-on collision. Suppose that the nearest PBH is placed at a distance $y > x$ from the center of mass of the pair at matter-radiation equality. Assuming a random distribution of PBHs, the probability of finding the relevant distances within the intervals $(x,x+dx)$ and $(y,y+dy)$ at matter-radiation equality is given by~\cite{Ioka:1998nz}
\begin{align}
P (x,y) \dd x \dd y & = n^2_\mrm{PBH}(4\pi x^2 \dd x) (4\pi y^2 \dd y)e^{-\frac{4\pi y^3}{3}n_{\text{PBH}}} \nonumber\\
& = \frac{9x^2y^2}{\overline{x}^6}\text{e}^{-y^3/\bar{x}^3} \dd x \dd y\,,
\end{align}
where $n_{\text{PBH}} = \rho_{\text{PBH}}(z_{\text{eq}})/M_{\text{PBH}}$ from Eq.~(\ref{rbara}) and  $\int_0^{\infty} dx \int_x^{\infty}dy P(x,y) = 1$. Since
only pairs satisfying the condition $x < f_{\text{PBH}}^{1/3}\overline{x}$ may form binaries, the fraction of formed binaries in the early Universe is estimated as 
\begin{equation}
P_{\text{bin}} = \int_0^{f_{\text{PBH}}^{1/3}\overline{x}}\dd x \int_x^{\infty}\dd y\, P(x,y) = 1-e^{-f_{\text{PBH}}}\,,   
\end{equation}
in good agreement with Ref.~\cite{Raidal:2017mfl} when the PBH clustering is neglected. 
For the case of our interest where only a small fraction of DM is in PBHs of masses
$(1\text{--}10)\,M_{\odot}$ PBHs, the fraction of isolated PBHs which acquire minihalos
is estimated as $1-P_{\text{bin}}=\text{exp}(-f_{\text{PBH}})\sim1$.

\section{Details of the Spherical Accretion Construction}
\label{app:accretion_details}

In this appendix we collect the derivations and implementation details underlying the spherical accretion model summarized in Sec.~\ref{sec.accretionmodel}. Our goal is to describe the formation of a QCD axion minihalo around an isolated PBH seed in a radiation--matter background, keeping explicit the assumptions that enter the shell initialization and subsequent nonlinear evolution.

\subsection{Equation of motion for a spherical shell}
\label{app:shell_eom}

We consider a spherically symmetric shell of physical radius \(r(t)\) enclosing a total mass
\begin{equation}
M_{\rm tot}(r)=M_{\rm PBH}+M_{\rm halo}(r)\,,
\end{equation}
where \(M_{\rm PBH}\) is the central PBH seed and \(M_{\rm halo}(r)\) denotes the effective collisionless matter mass enclosed by the shell. Neglecting dark energy during the epochs relevant for minihalo growth, the shell evolves in a background composed only of matter and radiation\,\footnote{The accretion process from the central PBH is cumulative; as time progresses, more shells
become detached from cosmological expansion and enter a virialized state. However, this accretion halts well before the current age of the Universe, specifically at the point when minihalos start to interact with nonlinear structures\,\cite{Berezinsky:2013fxa}. In the nonlinear regime, around redshifts $z \sim (30-10)$, the masses of minihalos reach their
maximum before being captured by galactic halos at
approximately $z \sim 6$\,\cite{CooraySheth2002}. Since the matter-dark energy
equality occurs at $z_{\text{m}\Lambda} = \mathcal{O}(0.1)$, we can reasonably
overlook the effects of dark energy and treat the Universe as being dominated by matter and radiation.}. 
Following the standard one-fluid secondary-infall treatment\,\cite{Kolb:1994fi}, we neglect the dynamical distinction between baryons and axion dark matter during halo formation and identify this effective collisionless component with the axion dark matter in the approximation $f_\chi = \Omega_{\text{DM}} / \Omega_{m} \simeq 1$. A fully coupled treatment of axion and baryonic components, including delayed baryonic infall after recombination, is beyond the scope of the present analysis. The physical axion fraction $f_\chi$ is subsequently restored in the normalization of the axion halo density.

The equation of motion for a shell may then be written as
\begin{equation}
\ddot r(t)
=
-\frac{G M_{\rm tot}}{r^2(t)}
-\frac{8\pi G}{3}\,\rho_r(t)\,r(t)\,,
\label{eq:app_newton_shell}
\end{equation}
where the radiation contribution is treated as homogeneous on the scales of interest. For each shell we choose the initial radius at horizon entry,
\begin{equation}
r_i \equiv r(t_i) = 2t_i\,,
\label{eq:app_ri}
\end{equation}
where \(t_i\) is the shell initial time and we have used $H^{-1}=2t$ within radiation domination. Since we are interested in minihalo formation around a pre-existing PBH in a regime where the axion already behaves as cold dark matter, we require
\begin{equation}
t_i \gtrsim \max(t_{\rm PBH},t_{\rm osc})\,,
\label{eq:app_ti_condition}
\end{equation}
with \(t_{\rm PBH}\) the PBH formation time and \(t_{\rm osc}\) the onset of axion oscillations. 

It is convenient to factor out the Hubble expansion and define the shell deviation variable \(b(t)\) through
\begin{equation}
r(t)=a(t)\,b(t)\,\xi\,,
\label{eq:app_rabxi}
\end{equation}
where \(\xi\) is the comoving shell radius. Pure background expansion corresponds to \(b(t)=1\). At the initial time, $r_i = a_ib_i\xi$, with $b_i \equiv b(t_i)$. We normalize the shell-deviation variable by setting $b_i =1$, which is equivalent to defining the comoving coordinate $\xi \equiv r_i/a_i$. The background scale factor satisfies the Friedmann equations
\begin{align}
\ddot a(t)
&=
-\frac{8\pi G}{3}\left[\frac{1}{2}\rho_m(t)+\rho_r(t)\right]a(t)\,,
\\[3pt]
H^2(t)
&=
\left(\frac{\dot a}{a}\right)^2
=
\frac{8\pi G}{3}\left[\rho_m(t)+\rho_r(t)\right]\,.
\end{align}
Passing to conformal time, \(\dd t = a \dd \eta\) and inserting Eq.~\eqref{eq:app_rabxi} into Eq.~\eqref{eq:app_newton_shell}, we obtain
\begin{align}
&a'(\eta)b'(\eta)+a(\eta)b''(\eta)\nonumber\\
&-\frac{4\pi G}{3}\,\rho_m(\eta)\,a^3(\eta)\,b(\eta)
+\frac{G M_{\rm tot}}{\xi^3 b^2(\eta)}
=0\,,
\label{eq:app_shell_eta}
\end{align}
where primes denote derivatives with respect to conformal time.

In a radiation--matter Universe, the scale factor is well approximated by\,\cite{Mukhanov:2005sc}
\begin{equation}
a(\eta)
=
a_{\rm eq}
\left[
\left(\alpha\frac{\eta}{\eta_{\rm eq}}\right)^2
+
2\left(\alpha\frac{\eta}{\eta_{\rm eq}}\right)
\right]\,,
\label{eq:app_aeta}
\end{equation}
where $\alpha\equiv \sqrt{2}-1$,  
$\eta_{\rm eq} = \alpha(\pi G \rho_{\rm eq} a^2_{\rm eq}/3)^{-1/2}$, \(a_{\rm eq}\equiv a(\eta_{\rm eq})\). At matter-radiation equality, we have $\rho_m(a_\text{eq}) = \rho_r(a_\text{eq})$, so that
\begin{equation}
\rho_{\rm eq}=2\rho_r(a_{\rm eq})=2\rho_m(a_{\rm eq})\,.
\end{equation}
Let the subscript \(i\) denote evaluation at the shell initial time \(t_i\). The initial total matter  density background is
\begin{equation}
\rho_{i}
=
\frac{\rho_{\rm eq}}{2}\left(\frac{a_{\rm eq}}{a_i}\right)^3\,.
\label{eq:app_rhoi}
\end{equation}
Since throughout this construction we condition on the presence of one heavy PBH at the origin, the heavy-PBH population is treated as a discrete source and is not included in the smooth background density that defines the shell overdensity. This is justified because the heavy-PBH fraction is assumed to satisfy \(f_{\rm PBH}\ll 1\) throughout our work. 

The total mass enclosed within the shell can therefore be parameterized as
\begin{align}
M_{\rm tot}(\xi)
&=
M_{\rm PBH}+M_{\rm halo}(\xi)\,,
\\[3pt]
&=
\frac{4\pi}{3}r_i^3(\xi)\,\rho_i\,[1+\delta_i(\xi)].
\label{eq:app_mtot_delta}
\end{align}
It follows that the initial density contrast is
\begin{equation}
\delta_i(\xi)=\delta_{\rm PBH}(r_i)+\delta_{{\rm halo},i}\,,
\label{eq:app_delta_split}
\end{equation}
with
\begin{equation}
\delta_{\rm PBH}(r_i)=\frac{3M_{\rm PBH}}{4\pi r_i^3\rho_i}\,,
\label{eq:app_deltaPBH}
\end{equation}
and
\begin{equation}
\delta_{{\rm halo},i}
=
\frac{3\left(M_{{\rm halo},i}-\bar M_{{\rm halo},i}\right)}{4\pi r_i^3\rho_i}\,.
\label{eq:app_deltahalo}
\end{equation}
Here, $\bar M_{{\rm halo},i}\equiv 4\pi r_i^3 \rho_i/3 $ denotes the smooth-background mass of the effective collisionless matter component enclosed within the same initial shell radius. The shell-crossing-free evolution assumed in the secondary-infall treatment implies mass conservation in the form
\begin{equation}
(1+\delta_i)b_i^3=(1+\delta(\eta,\xi))\,b^3(\eta)\,.
\label{eq:app_mass_conservation}
\end{equation}
When this relation is used to initialize the shell peculiar velocity, the time derivative is associated with the evolving smooth halo perturbation, while the central PBH mass held fixed.
 Introducing the dimensionless variable
\begin{equation}
y\equiv \frac{a(\eta)}{a_{\rm eq}}\,,
\label{eq:app_ydef}
\end{equation}
and using the standard scaling \(\rho_m(y)= (\rho_{\rm eq}/2)y^{-3}\), Eq.~\eqref{eq:app_shell_eta} reduces to
\begin{equation}
\begin{split}
(1+y)y\,b''(y)
&+
\left(1+\frac{3}{2}y\right)b'(y)\\
-
\frac{b(y)}{2}
&+
\frac{(1+\delta_i)b_i^3}{2b^2(y)}
=0\,.
\end{split}
\label{eq:app_shell_y}
\end{equation}
This is the shell equation quoted as Eq.\,\eqref{eq:secIV_main_shell} in the main text, which matches the same general structure in Ref.\,\cite{Kolb:1994fi}. However, the density contrast encodes the axion isocurvature perturbation, which is the differentiator term when we move from generic particle DM to the QCD axion.

\subsection{Window function and axion isocurvature variance}
\label{app:isocurvature_variance}

For the pre-inflationary QCD-axion scenario considered in Sec.~\ref{dominant}, we focus on the inflationary axion isocurvatute for shell initialization. 
The appropriate smoothing scale is the comoving shell radius,
\begin{equation}
R=\xi\,,
\end{equation}
and we adopt a normalized real-space top-hat window function\,\cite{Young:2024jsu},
\begin{equation}
W_R(\mathbf{x})
=
\left(\frac{4\pi R^3}{3}\right)^{-1}\Theta(R-|\mathbf{x}|)\,,
\label{eq:app_window_real}
\end{equation}
whose Fourier transform is
\begin{equation}
W(kR)
=
\frac{3}{(kR)^2}
\left[
\frac{\sin(kR)}{kR}-\cos(kR)
\right]\,.
\label{eq:app_window_fourier}
\end{equation}

Using the power-law parameterization of the primordial axion isocurvature spectrum introduced in Sec.~\ref{dominant},
\begin{equation}
\Delta_{\rm iso,a}^2(k)
=
\Delta_{\rm iso,a}^2(k_0)
\left(\frac{k}{k_0}\right)^{n_{\rm iso}-1}\,,
\end{equation}
with
\begin{equation}
\Delta_{\rm iso,a}^2(k_0)
\simeq
(f^{\rm DM}_a)^2\mathcal{A}^2_\text{anh}\frac{H_I^2}{\pi^2F_a^2\theta^2_{a,i}}\,,
\label{eq:app_deltaiso_pivot}
\end{equation}
we define the smoothed axion-isocurvature density contrast and the corresponding variance inside the initial shell as
\begin{align}
\delta_{\text{iso},R}(\mathbf{x},\eta_i) 
&\equiv \int \dd^3 \bv{x'} W_R(\mathbf{x}-\mathbf{x'})\delta_{\rm iso}(\mathbf{x'},\eta_i)\,,\\
\sigma^2_{\text{iso},i}(R)&\equiv \langle \delta^2_{\text{iso},R} (\mathbf{x},\eta_i)\rangle\,.
\end{align}

After the onset of coherent oscillations, axion density perturbations behave as pressureless cold dark matter on scales larger than the effective axion Jeans length, equivalently for comoving wavenumbers $k<k_J(a)$. For the QCD-axion masses and shell scales considered here, the modes that dominate the smoothed variance satisfy $k\sim R^{-1}\ll k_J(a_i)$. Axion gradient pressure is therefore negligible, and we may describe their subsequent evolution using the pressureless Mészáros growth factor\,\cite{Peacock:2003hh}
\begin{equation}
D_+(y)=1+\frac{3}{2}y\,,
\end{equation}
where $y \equiv a(\eta)/a_{\rm eq}$.
For this reason, in the regime $y_i\ll 1$ we use the factor $\left(1+(3/2)y_i\right)$ to describe the linear growth of the axion isocurvature density perturbation at shell initialization.

This prescription is valid only if the shell is initialized after the onset of axion oscillations, $t_i \gtrsim t_{\rm osc}$, so that the axion can be treated as pressureless matter, and only for subhorizon modes for which linear theory still applies. In our approach, the factor $D_+(y)$ is used only to determine the initial axion overdensity, whereas the subsequent shell evolution is computed with the full nonlinear spherical infall equation.

Therefore, we have
\begin{align}
\sigma_{{\rm iso},i}^2
&=D^2_+(y_i)
\int \frac{\dd^3\bv{k}}{(2\pi)^3}\,
\Delta_{\rm iso,a}^2(k)\,
W^2(kR)\,
\frac{2\pi^2}{k^3}
\nonumber\\[3pt]
&=
\left(1+\frac{3}{2}y_i\right)^2
\Delta_{\rm iso,a}^2(k_0)\nonumber\\
&\,\,\,\,\,\,\,\times \int_{R^{-1}}^\infty
\frac{\dd k}{k}
\left(\frac{k}{k_0}\right)^{n_{\rm iso}-1}\,W^2(kR)\,,
\label{eq:app_sigma_integral}
\end{align}
where we have isolated fluctuations with wavelengths no larger than the shell scale, treating modes with $k<R^{-1}$ as part of the locally coherent background and adopting the lower cutoff $k_{\rm min} = R^{-1}$.
Introducing \(x\equiv kR\), the integral factorizes as
\begin{equation}
\!\!\sigma_{{\rm iso},i}^2
=
\left(1+\frac{3}{2}y_i\right)^2
\Delta_{\rm iso,a}^2(k_0)
(Rk_0)^{1-n_{\rm iso}}\,I(n_{\rm iso})\,,
\label{eq:app_sigma_factored}
\end{equation}
where
\begin{equation}
I(n_{\rm iso})
\equiv
\int_1^\infty \dd x\,x^{n_{\rm iso}-2}\,\,W^2(x).
\label{eq:app_Iniso}
\end{equation}
Taking the positive square root, the axion-isocurvature rms amplitude used to initialize the shell is
\begin{equation}
\sigma_{{\rm iso},i}
=
\left(1+\frac{3}{2}y_i\right)
\Delta_{\rm iso,a}(k_0)(Rk_0)^{(1-n_{\rm iso})/2}
I^{1/2}(n_{\rm iso}).
\label{eq:app_sigma_iso_final}
\end{equation}
 In the spherical-infall calculation, we use a representative positive rms realization of the smooth perturbation, $\delta_{\text{halo},i}=+\sigma_{\text{iso},i}$, while the ensemble average of the isocurvature density contrast remains zero. 

For the benchmark scenario considered in this work, Sec.~\ref{dominant} gives \(n_{\rm iso}\simeq1\). We therefore set \(n_{\rm iso}=1\) in the numerical calculation, such that $(Rk_0)^{1-n_\text{iso}}=1$ and $I(n_\text{iso})=I(1)$, together with the Planck-saturated amplitude \(\Delta_{\rm iso,a}^2(k_0)=8.3\times10^{-11}\).

\subsection{Initial conditions for the shell evolution}
\label{app:initial_conditions}

The initial overdensity entering Eq.~\eqref{eq:app_shell_y} is the sum of the discrete PBH contribution and the smooth halo fluctuation.  
In order to isolate the additional smooth perturbation introduced by the pre-inflationary QCD-axion scenario, we model the initial halo perturbation by a representative positive rms axion-isocurvature fluctuation, allowing us to write 
\begin{equation}
\delta_i
=
\delta_{\rm PBH}(r_i)+\delta_{{\rm halo},i}
\simeq
\delta_{\rm PBH}(r_i)+\sigma_{{\rm iso},i}\,,
\label{eq:app_deltai_sigma}
\end{equation}
or explicitly
\begin{equation}
\delta_i \simeq \frac{3M_{\rm PBH}}{4\pi r_i^3\rho_i}
+
\sigma_{{\rm iso},i}\,.
\label{eq:app_deltai_explicit}
\end{equation}
We define the evolved density perturbation sourced by the central PBH for a fixed shell labeled by
$r_i$ (or equivalently $\xi$) as
\begin{equation}
\delta_{\rm PBH}(y;r_i)=\frac{3M_{\rm PBH}}{4\pi r^3(y)\rho_m(y)}\,,
\end{equation}
such that  $\delta_{\rm PBH}(y_i;r_i) = \delta_{\rm PBH}(r_i)$ at initialization.
Since $r(y)=a(y)b(y)\xi$ and $\rho_m(y) \propto a^{-3}(y)$, we have
\begin{equation}
\delta_{\rm PBH}(y;r_i)b^3(y) = \delta_{\rm PBH}(r_i)b_i^3\,,
\label{eq:massc}
\end{equation}
which implies that 
\begin{equation}
\frac{\dd \delta_\text{PBH}}{\dd y}= -3\delta_\text{PBH}(y;r_i)\frac{b'}{b}\,.
\label{eq:deltapbhdiff}
\end{equation}
To obtain the shell initial peculiar velocity, we differentiate the mass-conservation relation \eqref{eq:app_mass_conservation}. For a fixed shell, we write  
\begin{equation}
b(y)=
b_i(1+\delta_i)^{1/3}\,[1+\delta(y)]^{-1/3}\,,
\label{eq:masscc}
\end{equation}
where the total evolved perturbation is decomposed as
\begin{equation}
\delta(y)=
\delta_\text{PBH}(y;r_i)+
\delta_\text{halo}(y)\,,
\end{equation}
while at initialization we recover Eq.\,\eqref{eq:app_deltai_sigma}.
Differentiating Eq.\,\eqref{eq:masscc}, we have
\begin{equation}
\frac{b'}{b}= -\frac{\delta_\text{PBH}'+\delta_\text{halo}'}{3(1+\delta_\text{PBH}+\delta_\text{halo})}\,.
\end{equation}
Using Eq.\,\eqref{eq:deltapbhdiff}, the PBH contribution cancels out,  yielding 
\begin{equation}
(1+\delta_\text{halo})\frac{b'}{b} = -\frac{1}{3}\frac{\dd \delta_\text{halo}}{\dd y}\,.
\end{equation}
The linear isocurvature evolution derived above is used only to initialize the smooth halo perturbation. Therefore, at \(y=y_i\),
\begin{equation}
\delta_{\rm halo}(y_i)=\sigma_{{\rm iso},i},
\qquad
\left.
\frac{\dd \delta_{\rm halo}}{\dd y}
\right|_{y_i}
=
\left.
\frac{\dd \sigma_{\rm iso}}{\dd y}
\right|_{y_i}\,.
\end{equation}
Thus, the initial shell peculiar velocity takes the form
\begin{equation}
\left.\frac{\dd b}{\dd y}\right|_{y_i}
=
-\frac{1}{3(1+\sigma_{\text{iso},i})}
\left.\frac{\dd \sigma_{\text{iso}}}{\dd y}\right|_{y_i}\,,
\label{eq:app_bprime_general}
\end{equation}
where we have taken $b_i=1$ and 
\begin{equation}
\left.\frac{\dd \sigma_{\rm iso}}{\dd y}\right|_{y_i}
=
\frac{3}{2}\Delta_\text{iso,a}(k_0)
(Rk_0)^{(1-n_{\rm iso})/2}
I^{1/2}(n_{\rm iso})\,,
\label{eq:app_sigma_prime}
\end{equation}
from Eq.~\eqref{eq:app_sigma_iso_final}.

In the numerical implementation, each shell is initialized with
\begin{equation}
r_i=2t_i,
\,
b(y_i)=1,
\,
b'(y_i)=
-\frac{1}{3(1+\sigma_{{\rm iso},i})}
\left.\frac{\dd \sigma_{\rm iso}}{\dd y}\right|_{y_i},
\label{eq:app_num_init}
\end{equation}
subject to the consistency conditions
\begin{equation}
t_i \gtrsim \max(t_{\rm PBH},t_{\rm osc})\,,
\qquad
\sigma_{{\rm iso},i}\lesssim 0.1\,.
\label{eq:app_consistency}
\end{equation}
The purpose of the latter bound is to ensure that the smooth fluctuation used to seed the shell is still in the perturbative regime at the initial time, even though the later shell evolution is treated nonlinearly.

\subsection{Turnaround, virialization, and reconstructed density profile}
\label{app:turnaround}

Given the initial conditions above, Eq.~\eqref{eq:app_shell_y} is integrated until the shell reaches turnaround, defined by
\begin{equation}
\dot r(t_a)=0\,.
\end{equation}
Using \(r=a b \xi\), this condition becomes
\begin{equation}
b(y_{\max})+y_{\max}b'(y_{\max})=0\,,
\label{eq:app_turnaround}
\end{equation}
where \(y_{\max}\equiv a(t_a)/a_{\rm eq}\) and \(b_{\max}\equiv b(y_{\max})\).

We adopt the conventional approximation where the
virial radius is associated with the shell turnaround radius as\,\cite{CooraySheth2002}
\begin{equation}
R_{\rm halo}
=
\frac{r_{\max}}{2}
=
\frac{a_{\max}b_{\max}\xi}{2}\,,
\label{eq:app_Rhalo}
\end{equation}
with \(a_{\max}=a(t_a)\).  The mean enclosed axion density
takes the form
\begin{align}
\bar\rho_{\rm halo}(<R_{\rm halo})
 &=
\frac{6f_\chi M_{{\rm halo},i}(<r_i)}{\pi r_{\max}^3}\,,\\
& =
\frac{f_\chi\rho_{\rm eq}(1+\sigma_{{\rm iso},i})}{y_{\max}^3 b_{\max}^3/4}\,.
\label{eq:app_rhohalo2}
\end{align}

By scanning over shells labeled by different initial radii \(r_i\) or equivalently different \(t_i\), one reconstructs the mean enclosed density profile. The factor $f_\chi\equiv \Omega_{\rm DM}/\Omega_{\rm m}$  extracts the collisionless dark-matter fraction from the total matter density\,\cite{Bringmann:2011ut}. 

The local axion density is obtained by differentiating the enclosed mass as 
\begin{align}
\rho_{\rm halo}(R_{\rm halo}) &= \frac{1}{4\pi R^2_{\rm halo}}\frac{\dd M_{{\rm halo},i}(<R_{\rm halo})}{\dd R_{\rm halo}}\,,\\
&=\left(1-\frac{\gamma}{3}  \right)\bar\rho_{\rm halo}(<R_{\rm halo})\,,
\label{eq:rhonum2}
\end{align}
where we have used the no-shell-crossing assumption, $M_{{\rm halo},i}(<r_i)=M_{{\rm halo}}(<R_{\rm halo})$, and  assumed a power-law behavior for the mean enclosed axion density, e.g. $\bar\rho_{\rm axion}(<R_{\rm halo}) \propto R_{\rm halo}^{-\gamma}$\,.

\section{Detectability of gravitational wave signal}
\label{app:matched_gw}

In \cref{sec:gw_sig}, we showed that the presence of an ultracompact minihalo around a PBH leads to a dephasing of the GW merger signal compared to the vacuum case. For detectability, we must know whether this phase demodulation significantly affects the reconstructed signal-to-noise ratio when matching the incoming GW signal in the presence of the minihalo with the incorrect vacuum template.

Therefore, consider two GW signal templates in Fourier space: a vacuum template $\hat{h}_\mrm{vac}(f)$ and a DM minihalo template $\hat{h}_\mrm{DM}(f)$, where hats represent Fourier transforms. We may write them as~\cite{Maggiore:2007ulw}
\begin{equation}
    \begin{split}
    \hat{h}_\mrm{vac}(f) &= h_0 f^{-7/6} e^{i \varphi_\mrm{vac}(f)}\,,\\
    \hat{h}_\mrm{DM}(f) &= \hat{h}_\mrm{vac}(f) e^{i\Delta\varphi_\mrm{DM}(f)} \,,
    \end{split}
\end{equation}
where $h_0$ is the GW strain, $\varphi_\mrm{vac}(f)$ is the GW phase in vacuum, and $\Delta\varphi_\mrm{DM}(f)$ is the dephasing caused by the presence of the DM minihalo. In a matched filter analysis, as is typically done in GW experiments, statistical quantities make use of the inner product $(h_1,h_2)$ between two templates $h_1$ and $h_2$, defined by~\cite{Maggiore:2007ulw}
\begin{equation}
    (h_1, h_2) \equiv 4 \Re\left[\int_{0}^\infty \frac{\funop{\hat{h}_1^*(f)} \funop{\hat{h}_2(f)}}{S_{hh}(f)} \dd f\right]\,,
\end{equation}
where $S_{hh}(f)$ is the one-sided strain power spectral density. 

The optimal signal-to-noise ratio is achieved when the signal and template match. Supposing that the true signal stems from an NS-PBH merger in the presence of a UCMH, the optimal squared SNR is then proportional to the quantity $\rho^2 \equiv (h_\mrm{DM}, h_\mrm{DM})$. This is to be compared with the SNR when there is a mismatch between the template and the signal, for which we are instead interested in the quantity $\rho'^2 \equiv (h_\mrm{vac}, h_\mrm{DM})$. The ratio between the matched and mismatched inner products then gives us a measure of how much the reconstructed SNR deteriorates when we use the incorrect, vacuum-only template to reconstruct the GW signal. This is given by
\begin{equation}
    \mcal{R} \equiv  \left\lvert\frac{(h_\mrm{vac}, h_\mrm{DM})}{(h_\mrm{DM}, h_\mrm{DM})}\right\lvert  = \frac{\left\lvert\int_{0}^\infty \frac{f^{-7/3}}{S_{hh}(f)}\cos [\Delta\varphi(f)]\dd f\right\lvert}{\int_{0}^\infty \frac{f^{-7/3}}{S_{hh}(f)}\dd f}\,.
    \label{eq:matched_ratio}
\end{equation}

To compute this, we assume an observation conducted with the future LISA GW detector~\cite{LISA:2024hlh,LISA2017}. Admitting a frequency window of sensitivity of $f \in [0.1\,\mrm{mHz},1\,\mrm{Hz}]$, LISA is a prime detector with which to differentiate between the vacuum and minihalo scenarios since, as shown in \cref{fig:phase_gw}, the dephasing due to the UCMH grows for lower frequencies~\cite{LISA:2024hlh,LISA2017}. Refs.~\cite{Edwards:2019tzf,Eda:2014kra, Eda:2013gg} employed LISA in similar searches for the same reason. We take the strain noise power spectral density from Ref.~\cite{Robson:2018ifk}. The phase difference $\Delta\varphi$ appearing in \cref{eq:matched_ratio} can come from an arbitrary source.

\begin{figure}[t!]
    \centering
    \includegraphics[width=0.48\textwidth]{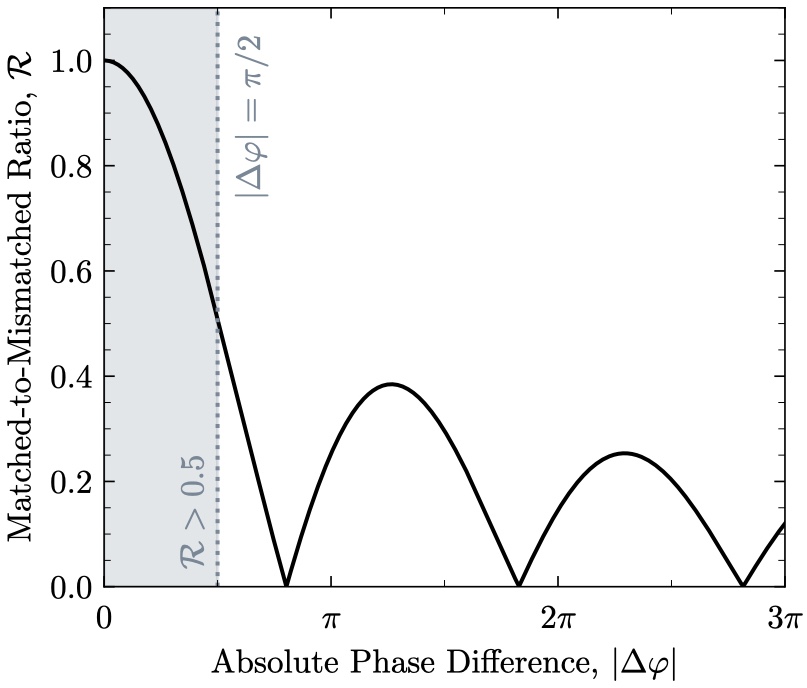}
    \caption{The ratio between the matched and mismatched inner products $\mcal{R}$ with absolute phase difference $|\Delta \varphi|$ when using the correct UCMH template and the incorrect vacuum template. When $|\Delta \varphi| < \pi/2$, the ratio becomes $\mcal{R} > 0.5$, and the SNR is not significantly deteriorated by the mismatch. We define this region to be where the UCMH is undetectable using gravitational waves alone, corresponding to the grey region in \cref{fig:phase_gw}.}
    \label{fig:matched_gw}
\end{figure}

We show how the matched-to-mismatched overlap ratio varies with increasing dephasing of the GW signal in \cref{fig:matched_gw}, serving as a proxy for how the SNR deteriorates. When $\mcal{R} \approx 1$, the reconstructed SNR is not significantly affected by the mismatch between signal and template, while $\mcal{R} = 0$ implies that the templates are orthogonal. We set a threshold of $\mcal{R} = 0.5$ to signify when the SNR should be significantly modified. For ratios worse than this, we claim that we can infer the presence of a UCMH surrounding the PBH in the PBH-NS merger. This threshold occurs when $|\Delta \varphi| = \pi/2$, corresponding to the grey filled regions in \cref{fig:matched_gw,fig:phase_gw}.

\section{Enhanced DM density at the conversion radius }
\label{App:EnhancedDM density}

We provide a detailed derivation of the DM density enhancement at the NS conversion radius. Following the local Liouville formalism used in Ref.\,\cite{Hook:2018iia} for a homogeneous DM background, we explicitly account for velocity-frame conventions and the impact-parameter measure. These elements are strictly necessary to treat our particular case of interest: an NS inspiral featuring a non-isotropic DM velocity distribution at asymptotic infinity.

In the PBH rest frame, let $\mathbf{w}$ and $\mathbf{v}_{\rm rel}$ denote the DM velocity and the NS velocity, respectively. In the NS rest frame, where we actually apply Liouville's theorem,
let $\textbf{V}_c$ be the DM velocity at the conversion point $\textbf{x}_c$ placed on the conversion surface. Similarly,  $\textbf{u}$ denotes the DM velocity at asymptotic infinity in the NS rest frame far away from the NS gravitational potential but still within the UCMH core.  

With this convention, the asymptotic DM velocity distributions in the NS and PBH rest frames are related by
\begin{equation}
f_\infty^\text{NS}(\textbf{u}) = f_\infty^\text{core}(\textbf{v}_\text{rel}+\textbf{u}) \,,
\end{equation}
where we have used 
$\textbf{u} = \textbf{w} - \textbf{v}_\text{rel}$.

Liouville's theorem tells us that
\begin{equation}
f_c^\text{NS}(\textbf{x}_c,\textbf{V}_c) = f_\infty^\text{NS}(\textbf{u}) \,,
\end{equation}
where $f_c^\text{NS}(\textbf{x}_c,\textbf{V}_c)$ is the density-normalized DM velocity distribution at the conversion point $\textbf{x}_c$ in the NS rest frame. As a result, the local DM density $\rho_c(\textbf{x}_c)$ is calculated from the expression
\begin{equation}
\begin{split}
\rho_c(\textbf{x}_c) &= \int \dd^3 \bv{V}_c f_c^\text{NS}(\textbf{x}_c,\textbf{V}_c) \\
&=
\int \dd^3\bv{V}_c f^\text{NS}_\infty(\textbf{u}(\textbf{x}_c,\textbf{V}_c))\,.
\end{split}
\label{eq:liouvilleapp}
\end{equation}
The mapping $\textbf{V}_c \rightarrow \textbf{u}$ is affected by gravitational focusing. For an isotropic Maxwell-Boltzmann distribution at infinity, as in the standard case treated by Ref.~\cite{Hook:2018iia}, the angular part of this map integrates out, and the density enhancement can be obtained from the speed relation alone. In the present case, however, the UCMH core distribution is anisotropic, so we keep track of the impact-parameter measure in order to justify a shell-averaged focusing factor.

We define a shell-averaged DM density in the spherical approximation as
\begin{align}
\bar \rho_c(r_c) &= \frac{1}{4\pi r_c^2}\int_{S_{r_c}} \dd A_c \rho_c(\textbf{x}_c) \nonumber\\ 
&  = \frac{1}{4\pi r_c^2}\int_{S_{r_c}} \dd A_c
\int \dd^3\bv{V}_c f^\text{NS}_\infty \textbf{(}\textbf{u}(\textbf{x}_c,\textbf{V}_c)\textbf{)}\,,
\label{eq:liouvilleapp2}
\end{align}
where $S_{r_c}$ is the sphere of radius $r_c$  in the NS magnetosphere\,\footnote{Note that  this is a spherical approximation to the true conversion surface. In the full magnetospheric problem, the conversion radius depends on angle and rotation phase, Eq.\,\eqref{eq:rcfull}. For our analytic estimate, we follow the same local radial-patch approximation applied in Ref.\,\cite{Hook:2018iia}}. The idea is to transform the local variables $(\textbf{x}_c, \textbf{V}_c)$ to the corresponding asymptotic set of variables.

The DM mass element crossing the area element $\dd A_c$ per unit time with local velocity element $\dd^3 \bv{V}_c$ at the sphere with radius $r=r_c$ takes the form 
\begin{align}
\dd \dot M_c =f_c^\text{NS}(\textbf{x}_c, \textbf{V}_c)
 |V_r| \dd A_c \dd^3\bv{V}_c \,,
\end{align}
%
where $V_r = \textbf{V}_c \cdot \hat{\textbf{r}}_c$ is the local radial velocity in the NS rest frame, with $\hat{\textbf{r}}_c = \textbf{x}_c/r_c$. Here, the absolute value appears because the mass flux through the conversion surface depends on the magnitude of the normal velocity.

\begin{figure}[t!]
    \centering
    \includegraphics[width=0.45\textwidth]{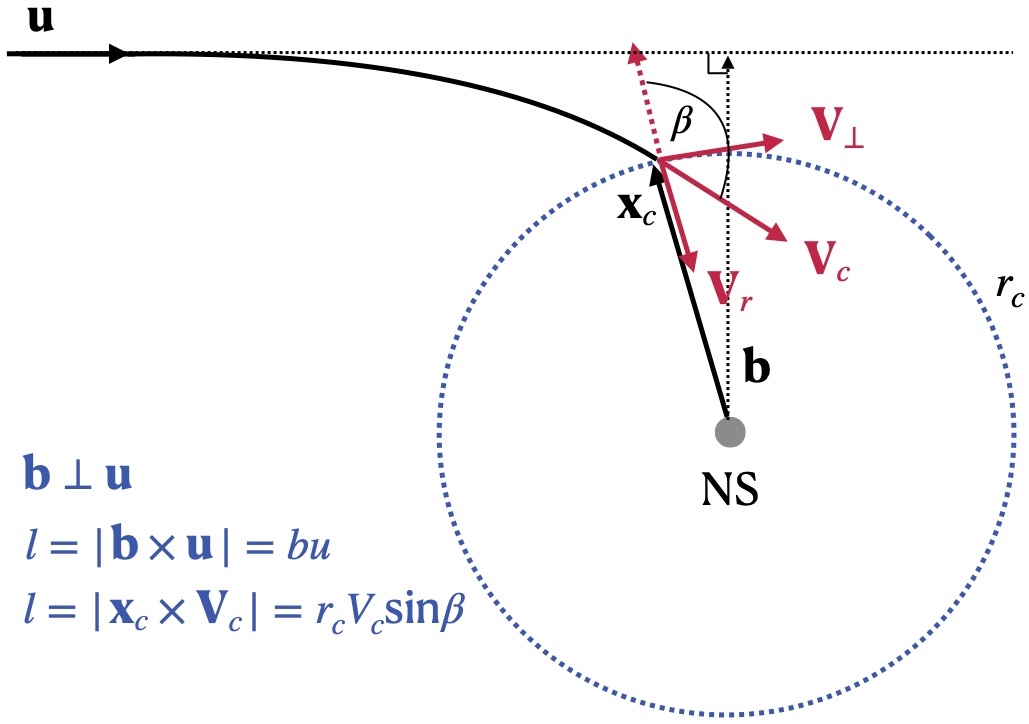}
      \includegraphics[width=0.45\textwidth]{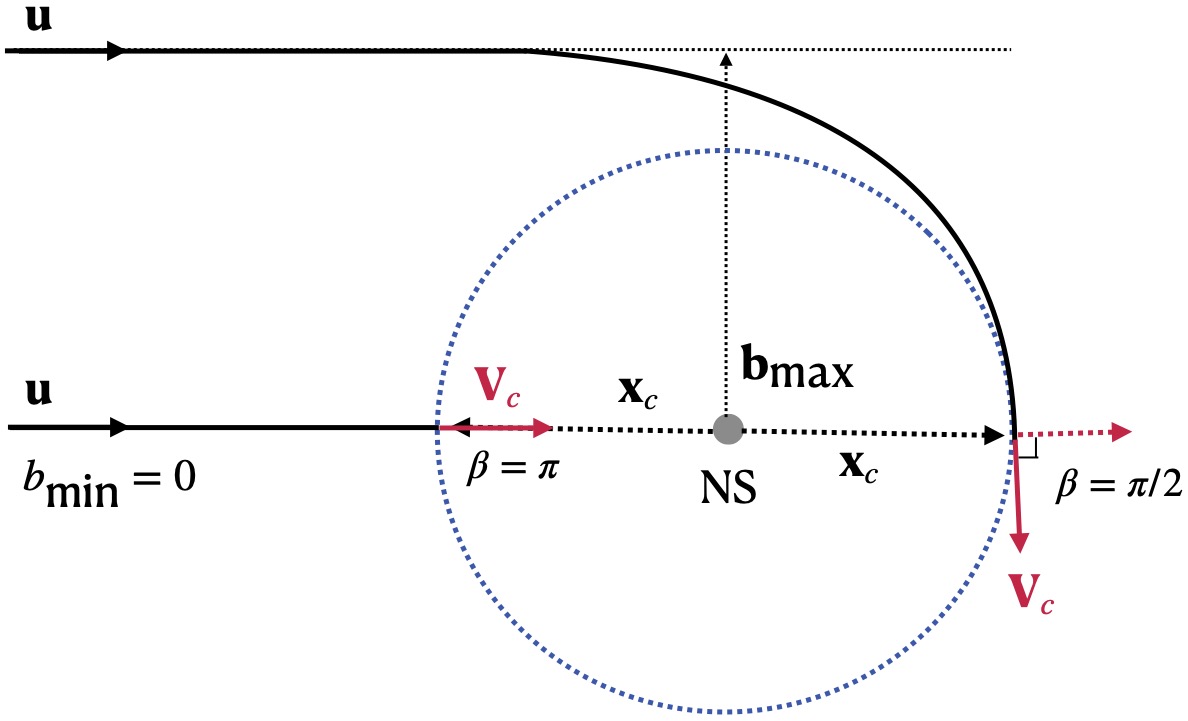}
    \caption{\textbf{Top:} Incoming DM particle with asymptotic velocity $\textbf{u}$ and impact parameter vector $\textbf{b}$, defined such that $\textbf{b} \perp \textbf{u}$. The particle crosses the conversion surface of radius $r_c$ at position $\textbf{x}_c$, with local velocity  $\textbf{V}_c$ forming an angle $\beta$ with $\textbf{x}_c$. \textbf{Bottom:} Limiting incoming trajectories corresponding to the  minimum and maximum impact parameters. For the grazing trajectory,  $b=b_\text{max}$ and $\hat{\textbf{V}}_c \cdot \hat{\textbf{x}}_c = 0$. For radial infall,  $b=b_\text{min}=0$ and $\hat{\textbf{V}}_c \cdot \hat{\textbf{x}}_c = -1$. }
    \label{fig:ip}
\end{figure}

Figure \ref{fig:ip} shows schematically an incoming DM particle to the NS crossing the conversion surface with radius $r_c$. The impact parameter lies in the plane perpendicular to the asymptotic velocity $\textbf{u}$. The impact-parameter area element reads as $\dd^2 b = b \dd b \dd \varphi$. The corresponding mass element of DM particles per unit of time  through the impact-parameter area element takes the form
\begin{equation}
\dd\dot M_\infty =f_\infty^\text{NS}(\textbf{u}(\textbf{x}_c, \textbf{V}_c)) \, u \dd^2b  \dd^3\bv{u}\,.
\end{equation}
Taking the DM flow as stationary and collisionless, we have $\dd \dot M_\infty  = \dd \dot M_c$ so that
\begin{equation}
\dd A_c \dd^3\bv{V}_c = \frac{u}{|V_r|} \dd^2b \dd^3\bv{u}\,,
\end{equation}
where we have used Liouville's theorem, $f_c^\text{NS}(\textbf{x}_c, \textbf{V}_c)=f_\infty^\text{NS}(\textbf{u})$. For an unbound trajectory crossing the shell, there are generally two branches, incoming and outgoing. If the conversion probability per crossing is small, which is the case of DM, both branches contribute to the phase-space density at the shell. This gives a net factor of two.

Replacing this measure expression into the shell average, Eq.\,\eqref{eq:liouvilleapp2}, we have
\begin{align}
\bar \rho_c(r_c) 
 &= \int \dd^3\bv{u}\,f^\text{NS}_\infty(\textbf{u})\nonumber\\ 
&\times \left[2\times\frac{1}{4\pi r_c^2}
\,\int_0^{2 \pi} \dd\varphi \int_{0}^{b_\text{max}} \dd b\,b  \frac{u}{|V_r|}\right]\,,
\label{eq:liouvilleapp3}
\end{align}
where the integration over the impact parameter is bounded from above by 
$b_{\text{max}}$, the maximum impact parameter for a particle that reaches the shell element. Now, we will deal first with the expression inside the bracket.
The conservation of energy for the incoming DM
in the NS potential reads as
\begin{align}
|\textbf{u}|^2 &= |\textbf{V}_c|^2 - 2\frac{GM_\text{NS}}{r_c}\,,\\
u^2 &= V_c^2 - v^2_\text{esc}\,,
\label{eq:u}
\end{align}
where $v^2_\text{esc}=2GM_\text{NS}/r_c$ is the escape velocity and $V_c = |\textbf{V}_c|$. On the other hand, the angular momentum conservation also gives extra confirmation.  At infinity and the conversion radius, its magnitude takes the form
\begin{align}
l &= |\textbf{b} \times \textbf{u}| = b\,u\,,\\
l & = |\textbf{x}_c \times \textbf{V}_c| =
r_c\, V_c\, \text{sin}\beta\,,
\end{align}
where $\beta$ is the angle between $\textbf{x}_c$ and $\textbf{V}_c$. Thus, $b\,u = r_c\, V_c\, \text{sin}\beta$ and the maximum impact parameter is 
\begin{equation}
b_\text{max}\equiv b(\beta=\pi/2)=\frac{r_c\, V_c}{u}\,.
\end{equation}
Since the local tangential speed relative to the NS-centered radial direction is just $V_\perp=b\,u/r_c$,
we can express $|V_r|$ in terms of the impact parameter as follows
\begin{align}
V_c^2 &= V_r^2 + V_\perp^2\,,\\
|V_r| &= V_c\sqrt{1-\left(\frac{b}{b_\text{max}}\right)^2}\,.
\label{eq:Vr}
\end{align}
Using Eqs.\,\eqref{eq:u} and \eqref{eq:Vr}, the square bracket is readily calculated as  $V_c/u$ so that
Eq.\,\eqref{eq:liouvilleapp3} takes the simpler form
\begin{equation}
\bar \rho_c(r_c) 
 = \int \dd^3\bv{u}\,f^\text{NS}_\infty(\textbf{u})\frac{\sqrt{u^2 + v_\text{esc}^2(r_c)}}{u}\,. 
\label{eq:liouvilleapp4}
\end{equation}
Using $f^\text{NS}_\infty(\textbf{u})=f^\text{core}_\infty(\textbf{v}_\text{rel}+\textbf{u})$, we have 
\begin{align}
&\bar\rho_c =  \frac{\rho_\text{core}}{(2\pi)^{3/2}\sigma_r \sigma_t^2}
\int_{-\infty}^{\infty}\dd u_r \int_{-\infty}^{\infty} \dd u_\theta
\int_{-\infty}^{\infty} \dd u_\phi\nonumber\\
&\times e^{\left[-\frac{u^2_r}{2\sigma^2_r} -\frac{(v_\text{rel}+u_\theta)^2+ u_\phi^2}{2\sigma^2_t}\right]}
\frac{\sqrt{u_r^2 +u_\theta^2+ u_\phi^2 + v_\text{esc}^2(r_c)}}{\sqrt{u_r^2 +u_\theta^2+ u_\phi^2}}\,.
\label{eq:rhocint}
\end{align}

To handle the angular integration we introduce polar coordinates in the tangential velocity plane, $u_\theta = u_\perp \text{cos}\varphi$ and $u_\phi = u_\perp \text{sin}\varphi$, so that
\begin{align}
\dd u_\theta \dd u_\phi &= u_\perp \dd u_\perp \dd u_\varphi\,,\\
(u_\theta + v_\text{rel})^2 + u^2_\phi&= u^2_\perp + v_\text{rel}^2 + 2 u_\perp v_\textbf{rel} \text{cos}\varphi\,.
\end{align}
The angular integral becomes a 
modified Bessel function of the first kind of zero order,
$\int_0^{2\pi}\dd \varphi e^{-(u_\perp v_\text{rel}\text{cos}\varphi)/(\sigma^2_t)} = 2\pi I_0(u_\perp v_\text{rel}/\sigma_t^2)$ and Eq.\,\eqref{eq:rhocint}  becomes
\begin{align}
&\bar\rho_c(r_c)
\simeq
\frac{\rho_{\rm core}}{\sqrt{2\pi}\sigma_r\sigma_t^2}e^{-\frac{v^2_\text{rel}}{2\sigma_t^2}}
\int_{-\infty}^\infty \dd u_r e^{-\frac{u^2_r}{2\sigma_r^2}}\times \nonumber\\
&\int_{0}^\infty \dd u_\perp u_\perp\,
e^{-\frac{u_\perp^2}{2\sigma_t^2}} I_0\left( \frac{u_\perp v_\text{rel}}{\sigma_t^2} \right)
\frac{\sqrt{u_r^2+u_\perp^2+v_{\rm esc}^2(r_c)}}{\sqrt{u_r^2+u_\perp^2}}\,.
\label{eq:rhoc_tangential_integral}
\end{align}
Since $\sigma_r \ll \sigma_t$, the radial Gaussian is narrow. In the limit $u_r \simeq 0$, we simply have $\int_{-\infty}^\infty \dd u_r e^{-u^2_r/(2\sigma_r^2)}=\sqrt{2\pi}\sigma_r$, so that Eq.\,\eqref{eq:rhoc_tangential_integral} takes the form
\begin{align}
\bar\rho_c(r_c)
\simeq
\frac{\rho_{\rm core}}{\sigma_t^2}&e^{-\frac{v^2_\text{rel}}{2\sigma_t^2}}
\int_0^\infty du_\perp\,
e^{-\frac{u_\perp^2}{2\sigma_t^2}}\times \nonumber\\
&I_0\left( \frac{u_\perp v_\text{rel}}{\sigma_t^2} \right)
\sqrt{u_\perp^2+v_{\rm esc}^2(r_c)}\,.
\end{align}

\section{Choice of radio integration time}
\label{sec:radio_integration_time}

Here we focus on the choice for the integration time used in the Green Bank Telescope sensitivity estimates.  We frame this calculation in a binary quasi-static snapshot regime and we treat the GBT integration time accordingly. Thus, we require the following hierarchy among the different timescales involved,
\begin{equation}
P \ll t_{\rm obs} \ll t_{\rm evol}\,,
\label{eq:tobs_hierarchy}
\end{equation}
where \(P\) is the neutron-star spin period and $t_{\rm evol} \equiv r/|\dot{r}|$ is the characteristic orbital-evolution time, where 
$\dot{r}$ obeys Eq.\,\eqref{eq:r_dot}.  Equation\,\eqref{eq:tobs_hierarchy} ensures the validity of the spin-averaged treatment for the flux density, Sec.\,\ref{subsec:SpectralFluxDensity}, and that the binary separation, orbital velocity, and signal bandwidth do not vary appreciably during one radio observation.

The total bandwidth entering the radio calculation is  given by Eq.\,\eqref{eq:main_Dnu_eff},
\begin{equation}
\Delta\nu^2
\simeq
\Delta\nu_{\rm int}^2
+
\Delta\nu_{\rm bulk}^2
+
\Delta\nu_{\rm corot}^2\,.
\label{eq:main_Dnu_eff2}
\end{equation}
for fixed axion mass and neutron-star properties, the binary radius dependence of the different broadening sources scale as  $\Delta\nu_\text{int}\propto r^{-2}$, Eqs.\,\eqref{eq:Dnuint_tangential}, $\Delta\nu_\text{bulk}\propto r^{-1/2}$, Eq. \eqref{eq:nubulk}, and $\Delta\nu_\text{corot}\simeq \text{constant}$. Therefore, the relative importance of the different broadening mechanisms changes as the binary evolves.

For an observational time much shorter than the characteristic orbital-evolution time, the fractional bandwidth drift during one snapshot can be estimated as
\begin{equation}
\frac{\delta\Delta\nu}{\Delta\nu}
\simeq
\frac{|\dot r|t_{\rm obs}}{r}\,
\left(\frac{
2\Delta\nu_{\rm int}^2+
\frac{1}{2}\Delta\nu_{\rm bulk}^2
}{
\Delta\nu^2
}\right).
\end{equation}
Requiring that the total bandwidth drift during one observation be at most about \(1\%\), e.g. $\delta \Delta\nu \lesssim 10^{-2}\Delta \nu$, the maximum integration time or, equivalently, observation time of a snapshot, takes the form
\begin{equation}
t_{\rm obs}\lesssim 
\frac{10^{-2} r}{|\dot r|}\,
\left(\frac{
\Delta\nu^2
}{
2\Delta\nu_{\rm int}^2+
\frac{1}{2}\Delta\nu_{\rm bulk}^2
}\right).
\end{equation}
For illustrative purposes, let us consider the fiducial Typical galactic NS benchmark of Table\, \ref{tab:NStargets}, with \(P=3\,{\rm s}\), \(B_0=10^{13}\,{\rm G}\), \(M_{\rm NS}=1.4M_\odot\), \(R_{\rm NS}=10\,{\rm km}\), and \(\theta_m=30^\circ\), and take \(m_a=10^{-5}\,{\rm eV}\), corresponding to \(\nu_{\rm peak}\simeq2.42\,{\rm GHz}\). We focus on the inspiral from
$r(f_{\rm GW}=0.1\,\text{Hz})$ to $r(f_{\rm GW}=1\,\text{Hz})$. In this case, as shown in Table\,\ref{tab:linewidth_evolution}, the bulk Doppler contribution dominates around the low-frequency end of the interval, whereas the intrinsic axion-dispersion width becomes comparable to (and eventually larger than) the bulk contribution as the binary approaches higher gravitational-wave frequencies (and smaller separation). The co-rotation contribution remains subdominant for this benchmark. Using the orbital evolution given in Eq.\,\eqref{eq:r_dot}, integrations of \((1-10)\) hr provide an excellent quasi-static approximation over the lower part of the frequency interval, whereas shorter snapshots are required toward the highest gravitational-wave frequencies, particularly for larger PBH masses. 
\begin{table}[H]
\centering
\caption{Characteristic radio linewidths and maximum quasi-static
integration time at the endpoints of the interval
$f_{\rm GW}=0.1$--$1\,{\rm Hz}$ for the fiducial Typical galactic NS
and $m_a=10^{-5}\,{\rm eV}$. The co-rotation contribution,
$\Delta\nu_{\rm corot}\sim0.1\,{\rm MHz}$, is subdominant for all
entries and is not shown.}
\label{tab:linewidth_evolution}

\scriptsize
\setlength{\tabcolsep}{3.0pt}
\renewcommand{\arraystretch}{1.15}
\vspace{0.3cm}
\begin{tabular}{cccccc}
\hline\hline
$M_{\rm PBH}$ &
$f_{\rm GW}$ &
$\Delta\nu_{\rm int}$ &
$\Delta\nu_{\rm bulk}$ &
$\Delta\nu$ &
$t_{\rm obs}^{\rm max}$ \\
&
[Hz] &
[MHz] &
[MHz] &
[MHz] &
[hr] \\
\hline
$1\,M_\odot$
& $0.1$ & $3.5$ & $16$  & $16$  & $417$ \\
& $1$   & $76$  & $34$  & $83$  & $3.9$ \\
\hline
$10\,M_\odot$
& $0.1$ & $18$  & $55$  & $58$  & $645$ \\
& $1$   & $385$ & $119$ & $403$ & $0.6$ \\
\hline\hline
\end{tabular}
\end{table}

\bibliography{main} 

\end{document}